\documentclass[journal]{IEEEtran}
\usepackage{amsmath,amsfonts}
\usepackage{algorithmic}
\usepackage{algorithm}
\usepackage{array}
\usepackage{textcomp}
\usepackage{stfloats}
\usepackage{url}
\usepackage{verbatim}
\usepackage{graphicx}
\usepackage{cite}
\usepackage[dvipsnames, svgnames, x11names]{xcolor}

\usepackage{colortbl}
\usepackage{color}
\usepackage{multirow}
\usepackage{makecell}
\usepackage{caption}
\usepackage{graphicx} 
\usepackage{float} 
\usepackage{subfigure} 
\usepackage{booktabs}
\usepackage{multirow}

\begin{document}

\title{CFAD: A Chinese Dataset for Fake Audio Detection}

\author{Haoxin Ma$^{1,2}$, Jiangyan Yi$^1$, Chenglong Wang$^1$, Xinrui Yan$^1$, Jianhua Tao$^{1,2,3}$
\\
\AND  Tao Wang$^{1,2}$, Shiming Wang$^1$,  Ruibo Fu$^1$ \\
$^1$NLPR, Institute of Automation, Chinese Academy of Sciences, China\\
$^2$School of Artificial Intelligence, University of Chinese Academy of Sciences, China \\
$^3$CAS Center for Excellence in Brain Science and Intelligence Technology, China \\
\texttt{\{haoxin.ma, jiangyan.yi\}@nlpr.ia.ac.cn} \\
}

\author{Haoxin Ma,~\IEEEmembership{Student Member,~IEEE}, Jiangyan Yi, ~\IEEEmembership{Member, ~IEEE},
Chenglong Wang,~\IEEEmembership{Student Member,~IEEE},
Xinrui Yan,~\IEEEmembership{Student Member,~IEEE},
Jianhua Tao,~\IEEEmembership{Senior Member,~IEEE},
Tao Wang,~\IEEEmembership{Student Member,~IEEE},
Shiming Wang,~\IEEEmembership{Student Member,~IEEE},
Ruibo Fu, ~\IEEEmembership{Member, ~IEEE} ,
\thanks{Haoxin Ma, Jiangyan Yi, Xinrui Yan, Jianhua Tao, Tao Wang, Le Xu and Ruibo Fu 
are with the National Laboratory of Pattern Recognition,
Institute of Automation, Chinese Academy of Sciences, Beijing
100190, China (e-mail:haoxin.ma@nlpr.ia.ac.cn; jiangyan.yi@nlpr.ia.ac.cn; yanxinrui2021@ia.ac.cn; jhtao@nlpr.ia.ac.cn; wangtao2018@ia.ac.cn; le.xu@nlpr.ia.ac.cn; ruibo.fu@nlpr.ia.ac.cn). Chenglong Wang and Shiming Wang is the Ph.D. candidate with the University of Science and Technology of China, Anhui, China (e-mail:chenglong@nlpr.ia.ac.cn; wsmzzz.mail@ustc.edu.cn). (Corresponding author: Jiangyan Yi and Jianhua Tao.).}
}

\markboth{Journal of \LaTeX\ Class Files,~Vol.~14, No.~8, August~2021}%
{Shell \MakeLowercase{\textit{et al.}}: A Sample Article Using IEEEtran.cls for IEEE Journals}



\maketitle

\begin{abstract}
Fake audio detection is a growing concern and some relevant datasets have been designed for research. However, there is no standard public Chinese dataset under complex conditions.
In this paper, we aim to fill in the gap and design a Chinese fake audio detection dataset (CFAD) for studying more generalized detection methods. Twelve mainstream speech-generation techniques are used to generate fake audio. To simulate the real-life scenarios, three noise datasets are selected for noise adding at five different signal-to-noise ratios, and six codecs are considered for audio transcoding (format conversion). CFAD dataset can be used not only for fake audio detection but also for detecting the algorithms of fake utterances for audio forensics. Baseline results are presented with analysis. The results that show fake audio detection methods with generalization remain challenging. The CFAD dataset is publicly available\footnote{https://zenodo.org/record/8122764}.
\end{abstract}

\begin{IEEEkeywords}
Fake audio detection, dataset, noise condition.
\end{IEEEkeywords}

\section{Introduction}
\IEEEPARstart{A}{dvanced} speech synthesis technology achieves a high level of naturalness, which brings a lot of convenience to our life. But the malicious use of generated speech can do us great harm. For example, the synthetic speech of public figures disseminated on social media can mislead the public's opinion. Criminals can use synthetic speech of a specific person to deceive another person or machine. Therefore, detecting fake audio is an urgent need and has attracted widespread attention in recent years. To carry out the relevant research, dataset construction is the foundation.


Early studies focus on the audio spoofing problem faced by ASV system and perform on private datasets \cite{lau2004vulnerability, lau2005testing,mariethoz2005can,zetterholm2007detection,hautamaki2013vectors, de2012synthetic, wu13_interspeech}. Until 2015, some public datasets\cite{wu2015sas,wu2015asvspoof,kinnunen2017asvspoof,wang2020asvspoof} are released, represented by ASVspoof databases. ASVspoof databases are used for a series of automatic speaker verification spoofing and countermeasures (ASVspoof) challenges, which have been held for 4 sessions so far. The release of a series of ASVspoof databases has greatly contributed to the advancement of anti-spoofing countermeasures.


Nowadays, people realize that fake audio attacks encountered in real-life scenarios are also worthy of attention. 
ASVspoof 2021 \cite{yamagishi2021asvspoof} extends the focus to the detection of deepfake speech in non-ASV scenarios. Besides, audio deepfake detection (ADD) challenges \cite{yi2022add, yi2023add} are held to address more issues posed by fake audio in real world and two sessions have taken place so far.
More datasets are constructed, such as FoR \cite{reimao2019dataset}, WaveFake \cite{frank2021wavefake}, HAD \cite{yi2021half}, and FMFCC-A \cite{zhang2022fmfcc} datasets.

These above-mentioned datasets facilitate the progress of fake audio detection. However, existing detection methods still lack generalizability\cite{muller2022does, luo2021capsule, gao2021generalized, chen2020generalization, pianese2022deepfake} and face difficulties when faced with out-of-domain or unseen situations. The fake audio in real life is not perfectly clean and often accompanied by disturbances such as noise or media codec. Moreover, most datasets are in English. Thus, there is a lack of a standard public dataset in the field of Chinese language. Although FMFCC-A is a public Chinese dataset, the fake speaker numbers (73 speakers), noise types (only Gaussian noise), and codec types (2 types) are not rich enough. Besides, no detailed labelling (e.g. noises, codecs) is provided in its uploaded dataset. ADD2022 and ADD2023 consider various real-world interfering factors, but they are inaccessible yet.

In this paper, we introduce a Chinese fake audio detection dataset, named CFAD. We hope it serves as a valuable complement to existing datasets and meets the following requirements: 1. It can evaluate the generalization of the model in unknown situations (including unknown types, unknown noises, and unknown codecs). 2. It can evaluate the robustness of the model under the complex conditions with interference of noise and codecs factors. 3. It provides detailed labels with specific operations, enabling researchers to customize the dataset according to their own experimental needs. Thus, CFAD dataset considers 12 types of fake audio, 11 of which are generated by different speech synthesis techniques and the remaining one is partially fake type\cite{yi2021half, zhang2021initial}. Partially fake audio is completely different from synthesis speech and thus can better evaluate the generalization of the detection model to unknown types. The real audio is collected from 6 different corpora to increase the diversity of real category distributions, which makes model less prone to artifact from a single database. For robustness evaluation, we additionally simulate background noise and media codecs that might occur in real life and provide detailed labels, including fake type, real source, noise type, signal noise ratio (SNR), and media codecs. Overall, CFAD dataset consists of three different versions, named clean, noisy, and codec versions. Relevant baseline experiments and analysis of results are presented. We hope the publication of the CFAD dataset can advance the progress in fake audio detection. 


The main contributions of our work are as follows: 

\begin{itemize}
    \item This is the first public Chinese standard dataset for fake audio detection under noisy conditions and transcoding (format conversion) conditions. We provide the detailed label of each audio to support flexible experimental settings for researchers.

    \item A variety of research related to fake audio detection is supported. For fake audio detection, the CFAD dataset can not only support the generalization studies on unseen types but also support the robustness studies under mismatched conditions. Besides, fake algorithm recognition studies for audio forensics can be conducted on the CFAD dataset. The corresponding baselines are provided to facilitate other researchers to compare against.
\end{itemize}

\section{Related Work}

\begin{table*}[t]
  \caption{Comparison between CFAD dataset versus prior works.}
  \label{tab:dataset_comparison}
  \centering
  \setlength\tabcolsep{3pt}
  \begin{tabular}{ccccccccc}
    \toprule
    Dataset & Language  & Condition  & Scenario &\# Speaker   &\# Utterance &  Fake & Real 
     & Accessibility\\
    \midrule  
    SAS & English &	Clean &	ASV	& \makecell[c]{Real: 106 \\Fake: 106} &
    \makecell[c]{More than \\ 652,615}
    &
    \makecell[l]{Types: VC,TTS \\Label: Yes}	&\makecell[l]{\#Resource: 1 \\ Label: Yes} &	Public  \\
    \midrule
    \makecell[c]{ASVspoof \\2015}  & English & Clean & ASV &\makecell[l]{Real: 106\\
    Fake: 106} &
    \makecell[c]{263,151}
    &
    \makecell[l]{Types: VC,TTS\\Label: Yes}&\makecell[l]{\#Resources: 1 \\Label: Yes} & Public\\
    \midrule
    \makecell[c]{Noisy \\Database} & English &Noisy &ASV &\makecell[l]{Real: 106\\Fake: 106} &
    \makecell[c]{About \\ 263,151}
    &
    \makecell[l]{Types: VC,TTS\\ Label: Yes} &\makecell[l]{\#Resources: 1\\
    Label: Yes} &	Restricted\\
    \midrule
    \makecell[c]{ASVspoof \\2017} &  English &Clean &	ASV	&\makecell[l]{Real: 42\\
Fake: 42} &
\makecell[c]{18,030}
    &\makecell[l]{Types: Replay\\
Label: Yes} &\makecell[l]{\#Resources: 1\\
Label: Yes} &	Public
 \\
 \midrule
    \makecell[c]{ASVspoof \\ 2019} & English &	Clean&	ASV	&\makecell[l]{Real: 107\\
Fake: 107}&
\makecell[c]{339,891}
    &
    \makecell[l]{Types: VC,TTS, \\ \quad \qquad Replay\\
Label: Yes}&\makecell[l]{\#Resources: 1\\
Label: Yes}&	Public
\\
\midrule
    FoR & English&	Clean&	Human&\makecell[l]{Real: 140\\
Fake: 33}& \makecell[c]{195,541}
    &\makecell[l]{Types: TTS\\
Label: No}&\makecell[l]{\#Resource: 4\\
Label: No}&	Public
\\
\midrule
    HAD& Chinese&	Clean&	Human&\makecell[l]{Real: 218\\
Fake: 218} &\makecell[c]{160,836}
    &\makecell[l]{Types: Partially fake\\
Label: No} &\makecell[l]{\#Resource: 1\\
Label: No}&	Restricted\\
\midrule
    WaveFake &\makecell[c]{English, \\
Japanese}&	Clean&	Human	&\makecell[l]{Real: 2\\
Fake: 2}&\makecell[c]{117,985}
    &\makecell[l]{Types: TTS\\
Label: Yes} &\makecell[l]{\#Resource: 1\\
Label: Yes}	&Public
\\
\midrule
\makecell[c]{ASVspoof \\ 2021} & English & \makecell[c]{Clean, \\Noisy, \\ Codec} & \makecell[c]{ASV,\\ Human} &
\makecell[l]{Real:133\\
Fake:133}&
\makecell[c]{1,566,273}
    &\makecell[l]{	Types: VC,TTS, \\ \quad \qquad Replay\\
Label: Yes}&\makecell[l]{\#Resource: 3 \\
Label: No}&	Public
\\
\midrule
\makecell[c]{FMFCC-A } & Chinese & \makecell[c]{Clean, \\Noisy.\\ Codec} & Human & \makecell[l]{Real: 58\\
Fake: 73} 
& 50,000 & \makecell[l]{Types: VC, TTS\\
Label: No} & \makecell[l]{\#Resource: 1\\
Label: No}& Public
\\
\midrule
\makecell[c]{ADD \\ 2022} & Chinese & \makecell[c]{Clean, \\Noisy, \\Codec} & Human &Unknown 
& 493,123 & \makecell[l]{Types: VC, TTS, \\ \quad \qquad Partially fake\\
Label: No} & \makecell[l]{\#Resource: 3\\
Label: No}& Restricted
\\
\midrule
\makecell[c]{ADD \\ 2023} & Chinese & \makecell[c]{Clean, \\Noisy.\\ Codec} & Human & Unknown & 517,068 & \makecell[l]{Types: TTS, \\ \quad \qquad Partially fake  \\
Label: Yes} & \makecell[l]{\#Resource: Unknown\\
Label: No}& Restricted
\\
\midrule
\textbf{Our CFAD} & Chinese & \makecell[c]{Clean, \\Noisy, \\Codec} &	Human &\makecell[l]{Real:1023\\
Fake:279}&\makecell[c]{347,400}
    &\makecell[l]{Types: TTS, \\ \quad \qquad Partially fake\\
Label: Yes}&\makecell[l]{\#Resource: 6\\
Label: Yes}&	Public \\
    \bottomrule
  \end{tabular}
\end{table*}

\begin{figure*}[t]
  \centering
    \includegraphics[width=15.5cm, height=9.5cm]{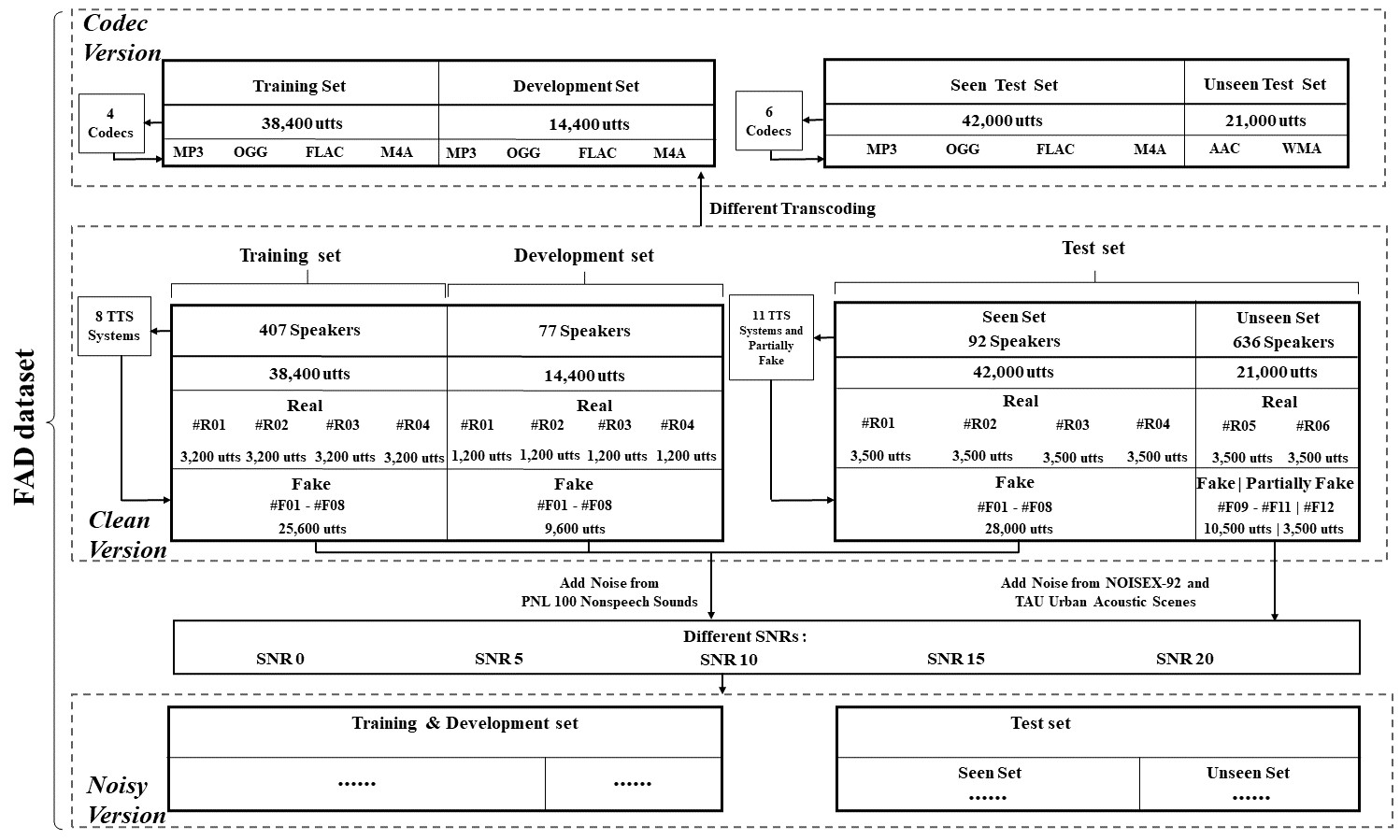}

  \caption{Partitions and construction of CFAD dataset. The middle part shows data partitions and details of clean version data, including the number of speakers, utterances, and types. The clean version serves as the foundation for the other two versions.
  The bottom part illustrates the process of adding noise to the clean data and creating the corresponding noisy version.
  The top part illustrates the process of audio transcoding to construct the corresponding codec version.
  }
    \label{fig:design}
\end{figure*}
In this section, we briefly introduce several other related fake audio detection datasets and then compare them with our CFAD dataset. We summarize the spoofing scenarios of the dataset into two categories: the ASV system and the human auditory system. In the human scenario, the main concern is fake speech in real life, especially those spread on social media, and replay audio is regarded as real audio.

For the datasets used to spoof the ASV system, SAS \cite{wu2015sas} corpus is published in 2015. The real audio is from a multi-speaker English
speech database, voice cloning toolkit (VCTK) \cite{veaux2017superseded}. The fake audio is generated by 2 speech synthesis systems and 7 voice conversion systems. All the spoofing systems are trained with VCTK. The background of audio is clean.
SAS is also used for supporting the ASVspoof 2015 \cite{wu2015asvspoof} but types of fake audio increase to 10.
Then, a noisy database \cite{tian2016spoofing} is built based on ASVspoof 2015 to investigate of spoofing detection under additive noisy conditions. It is generated by artificially adding five types of background noises at three different SNRs. But the database is unavailable yet.
ASVspoof 2017 (V2) \cite{delgado2018asvspoof} database is an improved version of the ASVspoof 2017 challenge, correcting some data anomalies. It focuses on replay attacks. The real audio of it is from RedDots \cite{lee2015reddots} corpus. The fake audio is collected by replaying and recording real utterances under more than 50 different configurations.
ASVspoof 2019 \cite{wang2020asvspoof} dataset is one of the most commonly used dataset in anti-spoof research. It contains two subsets: logical access (LA) and physical access (PA), collecting more diverse spoofing sources, still based on the VCTK database. 

ASVspoof 2021 \cite{liu2022asvspoof} database mainly focuses on the ASV system, but it adds a new task of speech deepfake which is designed to deceive the human auditory system. It uses the training and development partitions of the ASVspoof 2019 and releases three new evaluation partitions. The release of ASVspoof 2021 greatly promotes research in a variety of practical and complex scenarios. 

For the datasets used to spoof the human auditory system,
FoR \cite{reimao2019dataset} is a synthetic speech dataset. It contains fake audio from 7 open resources and real audio from 4 resources. 
HAD dataset \cite{yi2021half} is designed for partially fake audio detection. 
It's generated by manipulating the original utterances with genuine or synthesized audio segments, but unavailable yet.
WaveFake \cite{frank2021wavefake} is a deepfake dataset published in 2021. It collects 10 sample sets from 6 different network architectures across two languages (English and Japanese). The real audio is from LJSPEECH dataset \cite{ito2017lj} and JSUT dataset \cite{sonobe2017jsut}. FMFCC-A \cite{zhang2022fmfcc} is a Chinese synthetic speech detection dataset, containing 13 types of fake audio. It conducts noise addition and format conversion to test the performance of detection models. But, no detailed annotation is given on the label of each utterance. 
ADD 2022 \cite{yi2022add} dataset is used to support ADD 2022 challenge. It considers low-quality situations represented by background noises and partially fake types. But the specific generating method and the label of the test set have not been published. ADD 2023 \cite{yi2023add} further proposes the new tasks of manipulation region location and deepfake algorithm recognition, but still unpublished.

Table \ref{tab:dataset_comparison} highlights the differences. The \textit{Label} in the \textit{Fake} and \textit{Real} columns refers to whether the label of generation method or data source is given.
We can find that our CFAD dataset has various fake types and more real audio sources with various speakers. In addition to regular clean audio, our CFAD dataset also considers noisy conditions and audio transcoding(format conversion). In particular, our CFAD dataset is the only Chinese dataset that surpasses binary real/fake label and provides detailed information, including the source of real audio and the configuration of the post-processing operations (e.g. noises, SNRs, codecs).

\section{Dataset Design}
In this section, we first introduce our dataset design policy. Then we demonstrate the construction process of the dataset, including the resources of the clean real audio, the generation of the clean fake audio, the simulation process of the noisy audio, and the audio format conversion of each utterance. Finally, the overall statistics of the CFAD dataset are given.

\begin{table}[t] 
    \caption{The overall statistics of our CFAD dataset.\#Spk denotes the number of speakers. \#Utt denotes the total utterance number.}
    \label{tab:FADdataset}
    \centering
    \setlength\tabcolsep{3pt}
    \begin{tabular}{ c|cc|cc|cc }
    \toprule
    \multirow{2}*{Type} &  \multicolumn{2}{c|}{Clean} & \multicolumn{2}{c|}{Noisy} &  \multicolumn{2}{c}{Codec}\\
    \cline{2-7}
    ~  & \#Spk &\#Utt  &  \#Spk & \#Utt & \#Spk &\#Utt\\
    \midrule
     Train   & 407 &38400 & 407  & 38400 & 407 &38400\\
     Dev  &77 & 14400 &77&14400 &77 & 14400\\
     Seen Test  &92  & 42000 &92 &42000 &92  & 42000\\
     Unseen Test  &636 & 21000 &636 & 21000 &636 & 21000\\
     \cline{1-7}
     Total & 1212 & 115800 & 1212 & 115800 & 1212 & 115800\\
    \bottomrule
  \end{tabular}
\end{table}

\begin{table*}[t]
    \caption{The detail statistics of clean version of our CFAD dataset. 
    \#Spk denotes the number of speakes. \#Utt denotes the total utterance number.  \protect\\
    $^\ast$ Because the speakers in different types of fake audios overlap, the total number of speakers is not equal to sum of each type.}
    \label{tab:clean_dataset}
    \centering
    \setlength\tabcolsep{9pt}
    \begin{tabular}{ cccccccccc }
    \toprule
    \multirow{2}*{Label} & \multirow{2}*{Subset} & \multicolumn{2}{c}{Train} & \multicolumn{2}{c}{Dev}  &
    \multicolumn{2}{c}{Seen Test} & \multicolumn{2}{c}{Unseen Test}\\
    \cline{3-10}
    ~ & ~ & \#Utt & \#Spk & \#Utt & \#Spk
    & \#Utt & \#Spk & \#Utt & \#Spk\\
    \midrule
    \multirow{7}*{Real}& R01 &3200 &340 & 1200& 40 &3500 &20 &0 &0    \\
    ~ & R02 &  3200 &40 &1200 &20 &3500 &50 &0 &0    \\
    ~ & R03 &  3200 & 17 &1200 &14 &3500 &25  &0 &0    \\
    ~ & R04 &  3200 & 10 &1200 &3 &3500 & 7  &0 &0        \\
    ~ & R05 & 0 & 0 &0 &0 &0 &0 &  3500 &309        \\
    ~ & R06 & 0 & 0 &0 &0 &0 &0 &  3500 &138        \\
    ~ & Total & 12800& 407 &4800 &77 &14000 &92 &7000 &447         \\
    \midrule
    \multirow{13}*{Fake} & F01 & 3200 & 40 &1200 &20 &3500 &50 &0 &0   \\
    ~ & F02  & 3200 & 40 &1200 &13 &3500 &30 &0 &0     \\
    ~ & F03  & 3200 & 40 &1200 &13 &3500 &30 &0 &0     \\
    ~ & F04  & 3200 & 40 &1200 &20 &3500 &50 &0 &0    \\
    ~ & F05  & 3200 & 33 &1200 &13 &3500 &24 &0 &0    \\
    ~ & F06  & 3200 & 40 &1200 &9 &3500 &47 &0 &0     \\
    ~ & F07   & 3200 & 33 &1200 &13 &3500 &24 &0 &0    \\
    ~ & F08   & 3200 & 33 &1200 &13 &3500 &24 &0 &0    \\
    ~ & F09   & 0 & 0 &0 &0 & 0 & 0 &3500 &13   \\
    ~ & F10   & 0 & 0 &0 &0 & 0 & 0 &3500 &24   \\
    ~ & F11    & 0 & 0 &0 &0 & 0 & 0 &3500 &24   \\
    ~ & F12     & 0 & 0 &0 &0 &0 &0 &3500 &159  \\
    ~ & Total   &25600 & 40$^\ast$  &9600 & 20$^\ast$ & 28000 & 50$^\ast$ & 14000 & 189  \\
    \bottomrule
  \end{tabular}
\end{table*}

\begin{table}[t]
    \caption{The detail statistics of noisy version of our CFAD dataset.\#Utterances denotes the utterance number.}
    \label{tab:noisy_dataset}
    \centering
    \setlength\tabcolsep{5pt}
    \begin{tabular}{ cccccc }
    \toprule
    \multirow{2}*{Label} & \multirow{2}*{SNR} & \multicolumn{4}{c}{\#Utterances} \\
    \cline{3-6}
    ~ & ~ & Train & Dev & Seen Test & Unseen Test\\
    \midrule
    \multirow{6}*{Real} & 0dB& 3288 &	1019&	2503&	1062
        \\
    ~ & 5dB & 2073 &	813&	2512&	1312\\
    ~ & 10dB & 2072	&816&	2525&	1312
       \\
    ~ & 15dB & 2065	& 822&	2525&	1310
     \\
    ~ & 20dB & 3302	 &1330&	3935&	2004
      \\
    ~ & Total &12800 & 4800 & 14000 & 7000         \\
    \midrule
   \multirow{6}*{Fake} & 0dB &6006 &	1946&	4623&	2444
     \\
    ~ & 5dB &4164 &	1662&	5154&	2582
   \\
    ~ & 10dB &4165 &	1661&	5149&	2578
       \\
    ~ & 15dB &4161	 &1563&	5153&	2596
      \\
    ~ & 20dB &7104 &	2768&	7921&	3800
     \\
    ~ & Total & 25600  & 9600 & 28000&   14000     \\
    \bottomrule
  \end{tabular}
\end{table}

\begin{table}[t] 
    \caption{The detail statistics of codec version of our CFAD dataset.\#Utterances denotes the utterance number.}
    \label{tab:compressed_dataset}
    \centering
    \setlength\tabcolsep{4pt}
    \begin{tabular}{ cccccc }
    \toprule
    \multirow{2}*{Label} & \multirow{2}*{Codec} & \multicolumn{4}{c}{\#Utterances} \\
    \cline{3-6}
    ~ & ~ & Train & Dev & Seen Test & Unseen Test\\
    \midrule
    \multirow{6}*{Real} & Mp3 &3157 &1163 &3487 &0 
    \\
    ~ & Flac &3302 &1186 &3587 & 0 \\
    ~ & Ogg & 3231 &1201 & 3558& 0\\
    ~ & M4a & 3110 &1250 &3368 &0 \\
    ~ & Aac & 0 &0 & 0& 3496 \\
    ~ & Wma &  0 &0 & 0& 3504\\
    ~ & Total &12800 & 4800 & 14000 & 7000         \\
    \midrule
   \multirow{6}*{Fake} & Mp3 & 6362 & 2368 & 7030 &0
     \\
     ~ & Flac &6411 &2385 &7092  & 0 \\
    ~ & Ogg & 6405 &2407 &7045& 0\\
    ~ & M4a & 6422 &2440 &6833&0  \\
    ~ & Aac & 0 &0 &0 & 6915\\
    ~ & Wma & 0 &0 &0 & 7085\\
    ~ & Total &12800 & 4800 & 14000 & 7000         \\   
    \bottomrule
  \end{tabular}
\end{table}

\subsection{Design Policy}
The CFAD dataset is designed to evaluate the methods of fake audio detection and fake algorithms recognition and other relevant studies. 
The audio in real life is usually under noisy conditions. Besides audio on different social media adopts different codecs and storage formats, and faces the operation of format conversion. Therefore, we simulate the corresponding situations to better study the robustness of the detection models.
The total CFAD dataset consists of three versions: clean version, noisy version, and codec version. Each version of the dataset is divided into disjoint training, development, and test sets in the same way. There is no speaker overlap across these three subsets. 
Each test set is further divided into seen and unseen test sets. Unseen test sets can evaluate the generalization of the methods to unknown types. It is worth mentioning that both real audio and fake audio in the unseen test set are unknown to the model.

For the noisy speech part, we select three noise databases for simulation. Additive noises are added to each audio in the clean dataset at 5 different SNRs. The additive noises of the unseen test set and the remaining subsets come from different noise databases. 

For the codec speech part, we select six different codecs. Two of them are applied for unseen test set.

Figure \ref{fig:design} demonstrates the construction process.
Overall, the generation procedure of CFAD dataset consists of three steps:

1) \textbf{Building clean version.} Collecting clean real audio from different clean speech sources and generating clean fake audio by different techniques.

2) \textbf{Building noisy version based on clean version.} Adding noise signals to clean audio at various SNRs. The noise addition is done with all clean audio data.

3) \textbf{Building codec version based on clean version.} 
Transforming each clean audio into different codec formats and converting back to the original format.

\subsection{Clean Real Audio Collection}
Most of the previous fake audio detection datasets focus on the diversity of fake audio, ignoring the differences existing among real audio in real-life scenarios. Similar to FoR and ASVspoof2021 datasets, we collected real speech from multiple sources to reduce unintentional bias.

Our clean real audio is from two aspects: open resources and self-recorded data. Five speech resources from OpenSLR\footnote{http://www.openslr.org/12/} platform are selected. We also recorded some audio by ourselves. Researchers and students in our research group are involved in the data recording process. The reward is 100 RMB per person for 1 hour of recording. Each speaker signed the agreement. All the real data resources are:

\begin{itemize}
    \item \textbf{R01: AISHELL1} \cite{bu2017aishell}. This dataset contains more than 170 hours of speech data from 400 native Chinese speakers with different accents, recorded by high-fidelity microphones in a quiet indoor environment. It's used for Chinese speech recognition.
    
    \item \textbf{R02: AISHELL3} \cite{AISHELL-3_2020}. This dataset contains roughly 85 hours from 218 native Chinese speakers, supporting speech synthesis. The audios are high-fidelity.
    
    \item \textbf{R03: THCHS-30} \cite{THCHS30_2015}. This dataset contains more than 30 hours of speech data from 40 speakers recorded by single carbon microphones at silent office, supporting Chinese speech recognition studies.
    
    \item \textbf{R04: MAGICDATA Mandarin Chinese Read Speech Corpus}. This dataset contains 755 hours of scripted read speech data from 1080 native Chinese speakers with different accents. Most of the data is recorded by mobile phone and all the data is collected in a quiet indoor environment.
    
    \item \textbf{R05: MAGICDATA Mandarin Chinese Conversational Speech Corpus} \cite{yang2022open}. This dataset contains 180 hours of Chinese conversational speech data from 633 native Chinese speakers with different accents. All the data is recorded by mobile phone in a quiet indoor environment. We split long conversational speeches into short speeches according to scripts.

    \item \textbf{R06: Self-Recording Corpus}: This dataset contains about 60 hours of speech data from 200 native Chinese speakers, recorded by mobile phone in a quiet indoor environment. The ratio of male to female is around 1:1. 
    
\end{itemize}

Among these 6 corpora, R01-R04 are used for the training, development, and seen test sets. R05-R06 are used for the unseen test set. For R01-R04, we sample 7900 utterances from each corpus, of which 3200 utterances are used for training set, 1200 utterances are used for development set, and the remaining 3500 utterances are used for seen test set. For R05 and R06, we sample 3500 utterances from each for unseen test set.

\subsection{Clean Fake Audio Generation}
For text-to-speech (TTS) systems, the vocoder is the last step in generating speech, and it will leave more artifacts that are potential to be detected than the acoustic model. Yan et al.\cite{yan2022initial} research on vocoder fingerprints and demonstrate that they have different characteristics. From this point of view, we select 11 representative speech synthesis methods based on their vocoders. Wang et al. \cite{wang2022spoofed} also follow a similar approach by creating spoofed data using vocoders. The TTS systems are all trained and generated using AISHELL3 corpus. 
Besides, we collected another type, partially fake, to better evaluate the generalization of the detection methods. The partially fake audio is generated based on AISHELL1. These 12 types of fake audio are as follows:

\begin{itemize}
    \item \textbf{F01: STRAIGHT}.
    A traditional-vocoder-based system. This system uses STRAIGHT \cite{kawahara2006straight} vocoder. STRAIGHT is an extension of the classical channel vocoder that exploits the advantages of progress in information processing technologies and a new conceptualization of the role of repetitive structures in speech sounds.We implement it with the open-source code on GitHub\footnote{https://github.com/HidekiKawahara/legacy\_STRAIGHT.git}.
    
    \item \textbf{F02: Griffin-Lim}. 
     A traditional-vocoder-based system. This system uses Griffin-Lim \cite{perraudin2013fast} vocoder. Griffin-Lim uses a phase reconstruction method based on the redundancy of the short-time Fourier transform and promotes the consistency of a spectrogram. We implement the system using librosa\footnote{https://librosa.org/} tool.
    
    \item \textbf{F03: LPCNet}.
    A neural-vocoder-based system. This system generates audio by LPCNet \cite{valin2019lpcnet} vocoder. LPCNet is used to reduce the complexity of neural synthesis by using linear prediction. It makes it easier to deploy neural synthesis applications on lower-power devices. We implement it with the open-source code on GitHub\footnote{https://github.com/xiph/LPCNet.git}.
    
    \item \textbf{F04: WaveNet}. 
    A neural-vocoder-based system which generates audio by WaveNet\cite{oord2016wavenet} vocoder. This system models the conditional probability to generate the next sample in the audio waveform, given all previous samples and possibly additional parameters. We implement it with the open-source code on GitHub\footnote{https://github.com/r9y9/wavenet\_vocoder.git}.
    
    \item \textbf{F05: PWG}. 
    A neural-vocoder-based system. This system generates audio by Parallel WaveGAN (PWG) \cite{yamamoto2020parallel} vocoder. PWG is a distillation-free, fast, and small-footprint waveform generation method using a generative adversarial network. We implement it with the open-source code on GitHub\footnote{https://github.com/kan-bayashi/ParallelWaveGAN.git \label{url:pwg}}.
    
    \item \textbf{F06: HifiGAN}. 
    A neural-vocoder-based system. This system generates audio by HifiGAN \cite{kong2020hifi} vocoder. HifiGAN consists of one generator and two discriminators: multi-scale and multi-period discriminators, which are trained adversarially. We implement it with the open-source code on GitHub\textsuperscript{\ref{url:pwg}}.

    \item \textbf{F07: Multiband-MelGan}. 
     A neural-vocoder-based system. This system uses Multiband-MelGan \cite{yang2021multi} vocoder. The generator of Multiband-MelGan produces sub-band signals which are subsequently summed back to full-band signals as discriminator input. We implement it with the open-source code on GitHub\textsuperscript{\ref{url:pwg}}.
    
    \item \textbf{F08: Style-MelGAN}. A neural-vocoder-based system. This system applies a light-weight vocoder, Style-MelGAN \cite{mustafa2021stylemelgan}. Style-MelGAN allows synthesis of high-fidelity speech with low computational complexity. We implement it with the open-source code on GitHub\textsuperscript{\ref{url:pwg}}.
    
    \item \textbf{F09: WORLD}. 
    A traditional-vocoder-based system. This system uses WORLD \cite{morise2016world} vocoder. WORLD can estimate fundamental frequency (F0), periodicity and spectral envelope and also generate the speech similar to the input speech using only the estimated parameters. We implement it with the open-source code on GitHub\footnote{https://github.com/mmorise/World.git}.
    
    \item \textbf{F10: FastSpeech-HifiGAN}. 
    An end-to-end system which generates audio from input text sequence.
    The acoustic model fastspeech \cite{ren2020fastspeech} is used to generates mel-spectrum. The vocoder is HifiGAN used for waveform reconstruction. We implement it with the open-source code on GitHub\footnote{https://github.com/espnet/espnet \label{url:esp}}.
    
    \item \textbf{F11: Tacotron-HifiGAN}.
   An end-to-end system simililar to F10. But the acoustic model of F11 is Tacotron2 \cite{wang2017tacotron}. We implement it with the open-source code on GitHub\textsuperscript{\ref{url:esp}}.
    
    \item \textbf{F12: Partially Fake}: Partially fake audio is obtained by clipping and splicing. The generation way is similar to HAD dataset \cite{yi2021half}.
    For the generation process, we randomly replace one named entity $A$ in the real utterance with a different entity $B$. The entity $B$ is generated by an LPCNet-based text-to-speech system. When splicing real and fake segments, we adjust the volume of different segments to be consistent. Only one place in each utterance is fake audio. 
    For the majority of the partially fake type utterance, fake regions are hidden within the real audio, while a few are located at the beginning or end of the real audio.
    
\end{itemize}

Among the 11 speech synthesis methods, F01-F02, and F09 are traditional-vocoder-based systems. F03-F08,and F10-F11 are neural-vocoder-based systems. Thus, for unseen test set, we select one of the traditional vocoders, two neural vocoders, and a partially fake type. They are F09-F12. In each type in F01-F08, 3200 utterances are used for training set, 1200 utterances are used for development set, and 3500 utterances are used for seen test set. For each type in F09-F12, 3500 utterances are used for unseen test set.

Due to the scarcity of open source high-fidelity multi-speaker datasets for speech synthesis, our TTS systems are trained on the AISHELL3 (R02) dataset. To mitigate potential interference caused by speaker acoustic characteristics or transcript contents, we carefully check and select speakers and contents. First, we make sure that transcripts of each speaker are distinct from one another. Next, we select a total of 40 speakers with ID ranging from SSB0005 to SSB0686 for the training set. We select 20 speakers with ID ranging from SSB0693 to SSB0778 for the development set and select 50 speakers with ID ranging from SSB0780 to SSB1402 for the test set. Thus, speakers of different partitions in R02 and F01-F11 follow the above settings. However, because of technical issues, some certain speakers' speech was not generated, resulting in less than 40, 20, or 50 speakers for some fake types.


\subsection{Noisy Audio Simulation}

Noisy audio is designed to reduce the gap between ideal laboratory conditions and those to be expected in the wild. To simulate the real-life scenarios, we artificially sample the noise signals and add them to clean audio at 5 different SNRs, which are 0dB, 5dB, 10dB, 15dB, and 20dB. 

Additive noises are selected from three widely-used noise databases: PNL 100 Nonspeech Sounds \cite{hu2010tandem}, NOISEX-92 \cite{varga1993assessment}, and TAU Urban Acoustic Scenes \cite{Mesaros2018_DCASE}.
There are 20 kinds of nonspeech, environmental sounds in PNL 100 Nonspeech Sounds. NOISEX-92 contains 15 kinds of noisy environments, including conventional stationary noise (e.g., white, pink) and other scenario noise. TAU Urban Acoustic Scenes database supports the challenge of detection and classification of acoustic scenes and events (DCASE 2022) and contains 10 different acoustic scenes.

For the training, development, and seen test sets in noisy version, we randomly select noise signals from PNL 100 Nonspeech Sounds and add them to the clean speech with a random SNR in [0dB, 5dB, 10dB, 15dB, and 20dB].
For unseen test set of the noisy version, NOISEX-92 and TAU Urban Acoustic Scenes database are used for the same operations.

The generation of noisy audio in our CFAD dataset can be defined as Equation \ref{eq:add_noise}:

\begin{equation}
    y_{noisy}(t)=x_{clean}(t)+n_{noise}(t)
  \label{eq:add_noise}
\end{equation}

where \textcolor{blue}{$t$} denotes the time index. $y_{noisy}$ is referred to as our noisy audio after adding noise. $x_{clean}$ is an utterance of clean dataset. $n_{noise}(t)$ denotes a noise signal of noise database.

\subsection{Audio Transcoding}
Audio in social media comes in a variety of formats (codecs), often accompanied by audio transcoding, which introduces distortion. The Codec version aims to quantify the robustness of the methods under different format conversions. We select a total of six codecs. For the training, development, and seen test sets in codec version, mp3, flac, ogg, and m4a are used. For the unseen test set of the codec version, aac, and wma are used.
Audio transcoding operation is operated on the audio in the clean version. Each clean audio will be randomly transformed with one of the candidate codecs and converted back to original WAV files using ffmpeg\footnote{http://ffmpeg.org} toolkits.

\subsection{Statistics}

In each version (clean, noisy, and codec versions) of the CFAD dataset, there are 138400 utterances in training set, 
14400 utterances in development set, 
42000 utterances in seen test set, 
and 21000 utterances in unseen test set.
The overall statistics are demonstrated in the Tabel \ref{tab:FADdataset}.

Detailed statistics of the clean version are shown in the Tabel \ref{tab:clean_dataset}, which provides the number of speakers and the number of utterances in different subsets for both real and fake audio. In the same partitions (train/dev/seen test/unseen test), the speakers in R02 and F01-F11 overlap, so the total number of speakers is not equal to the simple addition of the terms.
Tabel \ref{tab:noisy_dataset} demonstrates statistics of each subset in the noisy version according to the SNRs. Tabel \ref{tab:compressed_dataset} presents statistics of each subset in the codec version.

\section{Baselines}

In this section, three baselines for fake audio detection and two baselines for fake algorithm recognition are provided and available on GitHub\footnote{https://github.com/ADDchallenge/CFAD}. All experiments are implemented in Python. The neural network models are trained with one GPU of GeForce RTX 2080. Before experiments, the evaluation metrics for each task are briefly introduced. Then, the experimental setup and results are presented. Based on these, we analyze the results.

\begin{table*}[t]
    \caption{The overall EERs (\%) of the model trained with different training data sets.}
    \label{tab:overall_eer}
    \centering
    \setlength\tabcolsep{2pt}
    \begin{tabular}{ c|c|c|c|c|c|c|c }
    \toprule
    \multirow{2}*{Train Set}  & \multirow{2}*{Model} &
    \multicolumn{6}{c}{Test Set} \\
    \cline{3-8}
    ~ & ~ & Clean Seen & Clean Unseen &  Noisy Seen & Noisy Unseen  & Codec Seen & Codec Unseen \\
    \midrule
    \multirow{3}*{Clean} & LFCC-GMM & 6.47 &31.90 &29.79 & 30.31  & 9.28 &32.58     \\
    ~ & LFCC-LCNN  &  1.26 & 26.56 & 20.14 & 33.77 & 23.30 & 40.18\\
    ~ & RawNet2  &   14.70 & 42.32 & 30.25 & 39.82 &3.94	&27.38
    \\
    \midrule
   \multirow{3}*{Noisy} & LFCC-GMM &    15.31 &33.48 & 19.80 & 31.71 & 12.76 & 32.87\\
    ~ & LFCC-LCNN  &    3.43 &24.01 & 6.88 &29.67 & 25.86 & 55.93\\
    ~ & RawNet2   &   23.71 &42.99 & 19.68 & 40.01 & 9.78	&39.04
    \\

    \midrule
   \multirow{3}*{Codec} &
   LFCC-GMM &   13.33 &34.64 & 32.36 & 32.09 &
   5.26 & 34.39\\
    ~ &
    LFCC-LCNN  &  15.22 &38.73 & 33.80 &45.16 &
    1.45 & 25.86\\
    ~ &RawNet2   &  6.58 &	31.08	& 22.44 & 40.95 &	6.02 &	31.12    \\  
    \bottomrule
  \end{tabular}
\end{table*}
\subsection{Evaluation Metrics}
\subsubsection{Equal Error Rate}
Equal error rate (EER) \cite{wu2015asvspoof} is the metric for deepfake audio detection\cite{yi2022add,yamagishi2021asvspoof}. Let $P_{fa}(\theta)$ and $P_{miss}(\theta)$ denote the false alarm and miss rates at threshold $\theta$:
\begin{equation}
\label{eq:far}
    P_{fa}(\theta ) =\frac{\# \{fake \ trials \ with \ score >   \theta \}  }{\# \{total \ fake \ trials\}} 
\end{equation}

\begin{equation}
\label{eq:miss}
    P_{miss}(\theta ) =\frac{\# \{real \ trials \ with \ score \le   \theta \}  }{\# \{total \ real \ trials\}} 
\end{equation}
EER corresponds to the threshold $\theta_{EER}$ at which the two detection error rates are equal, i.e. $ EER=P_{fa}(\theta_{EER} )= P_{miss}(\theta_{EER} )$.
The lower the value of EER, the better performance of the model. 

\subsubsection{$F_1$-Score}

$F_1$-score is the metric for fake algorithm recognition \cite{yan2022initial, yan2022system}. Let $TP$, $FP$, and $FN$ denote the true positive, false positive,
and false negative, respectively. Precision and Recall can be calculated as follows:

\begin{equation}
\label{eq:precision}
   Precision =\frac{TP}{TP+FP} , \quad Recall =\frac{TP}{TP+FN} 
\end{equation}

$F_1$-score is given by the harmonic mean between Precision and Recall:
\begin{equation}
\label{eq:f1}
  F_1-score =2 \times \frac{Precision \times Recall}{Precision+Recall} 
\end{equation}

The higher the value of $F_1$-score, the better performance of the model. 

\subsection{Fake Audio Detection}
\begin{table*}[tb]
    \caption{The EERs (\%) of the model trained with different training data sets on different fake types.}
    \label{tab:fake_type}
    \centering
    \setlength\tabcolsep{3pt}
    \begin{tabular}{ cc|c|c|c|c|c|c|c|c|c|c|c|c }
    \toprule
    \multirow{2}*{Train Set} &  \multirow{2}*{Model} & \multicolumn{8}{c|}{Clean Seen Test} 
    & \multicolumn{4}{c}{Clean Unseen Test}
    \\
    \cline{3-14}
    ~ & ~ & F01 & F02 & F03 & F04 & F05 & F06 & F07 & F08  & F09 & F10 & F11 & F12 
    \\ 
    \midrule
    \multirow{3}*{Clean} &LFCC-GMM   &
    3.20 &3.30& 11.23 &10.60 &4.51 &5.32 &5.08 &3.00  &
    40.50 &12.46 &4.91 &49.43 \\
    ~ & LFCC-LCNN     &
    2.04& 2.34 &1.56 &0.50 &0.398 & 0.67 & 0.74 &0.32  &
    16.54 &7.73 &2.74 &70.63 \\
    ~ & RawNet2  & 13.04& 19.07 &10.15 &8.11 &24.77 &9.95 & 12.05& 16.77 &41.20 &46.98 &27.62 &51.61        \\
    \midrule
    \multirow{3}*{Noisy} & LFCC-GMM   & 21.72 & 10.26&	17.71&	22.03&	7.89&	10.62&	7.99&	4.28 &
    48.11 &	14.31 &5.71 &	44.31 
    \\
    ~ & LFCC-LCNN     & 5.46 &4.63 &3.56 &2.66 &1.96 &2.11 &2.90 & 0.64 &
    15.97 &3.18 &4.91 &59.11 
    \\
    ~ & RawNet2  & 25.62 &  35.58 &28.97 & 21.37 & 18.11 &21.40 &16.60 &13.56  &
    45.77 &36.60 &24.44 &57.41 \\
    \midrule
    \multirow{3}*{Codec} & 
     LFCC-GMM   & 9.90 &9.06 &19.09 & 16.82 & 12.00 & 12.32 &12.62 & 8.28 & 45.74 & 14.28 &5.57 &52.37
    \\
    ~& 
    LFCC-LCNN     & 3.48 &	4.06&2.29	&11.91	&18.87	&14.56&	36.81	&1.41&	8.83&	20.11&	46.37	&67.99
    \\
    ~ &  
    RawNet2  & 4.17	&3.42	&5.80 &	4.65&	2.51&	2.51	&2.64&	2.22	&26.41	&20.78	&12.61&	68.25
 \\
    \midrule
    \multirow{2}*{ } &  \multirow{2}*{ } & \multicolumn{8}{c|}{Noisy Seen Test} 
    & \multicolumn{4}{c}{Noisy Unseen Test}
    \\
    \cline{3-14}
    ~ & ~ & F01 & F02 & F03 & F04 & F05 & F06 & F07 & F08  & F09 & F10 & F11 & F12 
    \\ 
    \midrule
    \multirow{3}*{Clean} &LFCC-GMM   &  
    28.51 &21.08 &42.55 &29.13 &28.10 &22.71 &28.38 &33.91  &
    36.62 &23.52 &13.35& 36.25 \\
    ~ & LFCC-LCNN     & 
    22.70 &17.87 &27.96 &12.93 &19.17 &12.52 &18.52 &26.06 &
    38.77 &19.56 &11.74 &67.76 \\
    ~ & RawNet2 &  
    39.54 &40.34 &25.85 &24.50 &35.21 &21.33 &20.01 &29.35  &
    43.55 &41.77 &29.46 &42.66 \\
    \midrule
    \multirow{3}*{Noisy} & LFCC-GMM   & 
    35.79 &38.10 &39.10 &10.98 &5.05 &5.03 &5.25 &4.87  &
    39.96 &20.65 &15.73 &43.29\\
    ~ & LFCC-LCNN     &  
    9.73 &6.26 &13.10 &6.14 &4.61 &3.93 &5.08 &4.90  &
    29.86 &5.62 &10.33 &65.44 \\
    ~ & RawNet2          &
    20.67 &25.93 &28.15 & 18.35 &14.86 &17.56 &13.44 &11.02  &45.65 &33.34 &24.29 &49.94 
    \\
    \midrule
    \multirow{3}*{Codec} & 
     LFCC-GMM   & 28.71&	22.60	&44.17	&29.36	&33.81	&26.18	&33.55	&37.86	&38.83	&21.25	&16.30	&43.55
    \\
    ~&  LFCC-LCNN     &26.70	&16.34	&30.58	&30.80	&46.52	&28.36	&54.26	&35.98	&28.38	&33.12	&54.46	&61.99
    \\
    ~ &  RawNet2  &  11.02&	10.25	&18.43	&12.14&	2.27	&5.77&	4.62	&2.14	&51.29&	25.46	&30.11	&57.83
    \\
    \midrule
    \multirow{2}*{ } &  \multirow{2}*{ } & \multicolumn{8}{c|}{Codec Seen Test} 
    & \multicolumn{4}{c}{Codec Unseen Test}
    \\
    \cline{3-14}
    ~ & ~ & F01 & F02 & F03 & F04 & F05 & F06 & F07 & F08  & F09 & F10 & F11 & F12 
    \\ 
    \midrule
    \multirow{3}*{Clean} &
     LFCC-GMM   &  4.09&	4.11&	18.75&	14.34&	5.14&	7.94&	5.69&	3.37	&40.74	&13.89&	5.36	&51.29
     \\
    ~ & LFCC-LCNN     & 3.91 &	4.39	 &6.20	 &10.37 &	34.69 &	23.37	 &40.96	 &32.58	 &23.17	 &17.03 &	53.16	 &62.14
     \\
    ~ &  RawNet2 &  5.97	& 4.01	& 11.85	& 5.97	& 5.54	& 1.82& 	3.98& 	3.23& 	26.41	& 20.78	& 12.61	& 68.25
     \\
    \midrule
    \multirow{3}*{Noisy} & 
     LFCC-GMM   & 18.38  &	7.43  &	15.79  &	21.74  &	5.48  &	10.69	  &5.37	  &3.00	  &43.97	  &16.21  &	7.00   &45.03
    \\
    ~ &  LFCC-LCNN    &  9.51	&8.37	&9.39	&19.83	&37.62	&14.40	&39.09	&44.63	&21.43	&22.53	&37.90	&59.51
     \\
    ~ &  RawNet2   &29.25	&18.21	&36.33	&22.22	&18.66	&10.47	&18.90 &	16.80 	&51.29	&25.46	&30.11	&57.83
    \\
    \midrule
    \multirow{3}*{Codec} & 
     LFCC-GMM   & 3.21	&2.81	&9.57&	8.58&	3.28	&5.46	&3.69	&2.22	&42.83	&15.34&	5.74&	53.20
    \\
    ~&  LFCC-LCNN     &1.69	 &1.99	 &1.41 &	1.23 &	1.45	 &1.97 &	0.97 &	0.37 &	14.51	 &12.49	 &11.91 &	64.19
    \\
    ~ &  RawNet2  &  5.88&	3.93	&10.8&	5.9	&5.48	&1.79	&3.85	&3.02&	27.71&	18.52	&12.75	&69.88
    \\
    \bottomrule
  \end{tabular}
\end{table*}

In real-world scenarios, open-source TTS tools are easily accessible and can be used to generate fake audio. Those fake audio can quickly spread on social media and the Internet for unethical purposes. To address these threats, fake audio detection is of great significance. It aims to detect whether the input audio is real or fake.

Motivated by the baseline systems in ASVspoof challenges and ADD challenges, we choose three of them for fake audio detection task.
They are:

\begin{enumerate}
    \item \textbf{LFCC-GMM}: This is a Gaussian-mixture-model-based (GMM) system operating on linear frequency cepstral coefficients (LFCCs) \cite{lei2009mel}, which is the same as ASVspoof 2021\footnote{https://github.com/asvspoof-challenge/2021}. 
    
    \item \textbf{LFCC-LCNN}: This system operates upon LFCC features with a light convolutional neural network (LCNN). Unlike LFCC-GMM, the frame length and shift of LFCC are set to 20ms and 10ms respectively. LCNN model refers to \cite{lavrentyeva2019stc}, but the 28th layer adopts AdaptiveMaxPool2d.
    
    \item \textbf{RawNet2}: This is a full end-to-end system \cite{jung2020improved} that operates directly upon raw audio waveforms. It consists of sinc filters, 6 residual blocks followed by gated recurrent units (GRU), and a fully connected layer. 
    
\end{enumerate}

\begin{table*}[tb]
    \caption{The EERs (\%) of the model trained with different training data sets at different SNRs.}
    \label{tab:fake_snr}
    \centering
    \setlength\tabcolsep{5pt}
    \begin{tabular}{ cc|c|c|c|c|c|c|c|c|c|c}
    \toprule
    \multirow{2}*{Train Set} &  \multirow{2}*{Model} & \multicolumn{5}{c|}{Noisy Seen Test} 
    & \multicolumn{5}{c}{Noisy Unseen Test} \\
    \cline{3-12}
    ~ & ~ & 0dB & 5dB & 10dB & 15dB & 20dB  & 0dB & 5dB & 10dB & 15dB & 20dB \\ 
    \midrule
    \multirow{3}*{Clean} &LFCC-GMM   & 
    35.34 &33.83& 30.51 &27.25 &24.62 & 
    34.66 & 30.29 &29.23 &27.37 &29.34\\
    ~ & LFCC-LCNN     & 
    32.06 &27.00 &20.96 &15.59 &10.82 & 
    40.18 &39.12 & 35.13 &33.33 &30.29 \\
    ~ & RawNet2 &  
    27.38 &27.53 &29.09 &30.75 &31.18 & 
    44.25 & 39.42 &38.12 &38.87 &38.47  \\
    \midrule
    \multirow{3}*{Noisy} & LFCC-GMM   &  
    22.39 & 20.93 &19.83 &19.38 &17.93 &
    36.35 &33.03 &30.37 &30.27 & 28.64 \\
    ~ & LFCC-LCNN     & 
    10.59 & 9.49 &6.63 &5.21 &4.41 &
    32.39 & 33.65 &30.84 &29.67 &27.05     \\
    ~ & RawNet2 &
    19.90 & 19.77 &19.11 &18.92 &20.17  &
    39.26 &39.12 &39.77 &39.24 &41.13 \\
    \bottomrule
  \end{tabular}
\end{table*}

\begin{table}[tb] 
    \caption{The EERs (\%) of the model trained with different training data sets tested on different codecs.}
    \label{tab:fake_codec}
    \centering
    \setlength\tabcolsep{2pt}
    \begin{tabular}{ cc|c|c|c|c|c|c}
    \toprule
    \multirow{2}*{Train Set} &  \multirow{2}*{Model} & \multicolumn{4}{c|}{Codec Seen Test} & \multicolumn{2}{c}{Codec Unseen Test} \\
    \cline{3-8}
    ~ & ~ & Mp3 & Flac & Ogg & M4a &  Aac  &  Wma \\ 
    \midrule
    \multirow{3}*{Clean} &LFCC-GMM   & 13.64&	7.28 &	7.03&	9.28	&33.09&	32.10
    \\
    ~ & LFCC-LCNN     & 25.58 &	20.19 &	20.17	 &21.11	 &42.95	 &38.10
     \\
    ~ & RawNet2 & 3.69	& 3.90 & 	4.48& 	3.50 & 	27.37	& 27.42
    \\
    \midrule
    \multirow{3}*{Codec} & LFCC-GMM   &  5.49 &5.10	&4.92	&5.75	&35.26	&33.55
    \\
    ~ & LFCC-LCNN     & 1.95	&1.14	&1.32&	1.45	&24.95	&28.13
    \\
    ~ & RawNet2 &4.64&	6.41	&6.74	&6.29&	30.97	&31.17
     \\
    \bottomrule
  \end{tabular}
\end{table}

We conduct several groups of experiments to evaluate the
performance of baseline systems on different test sets of our CFAD dataset. 
For each version of the dataset, we train the model using only the respective training data and use the respective development data to select the best model.

Table \ref{tab:overall_eer} presents the results tested on a complete seen or unseen test set. 
In cases of data version matched (i.e., the model is trained and tested on the same version of the data, but the test set can either be seen or unseen), the LFCC-LCNN model achieves the best performance. Specifically, when trained on the clean data, the model achieves an EER of 1.26\% and 26.56\% on the clean seen and unseen test sets, respectively. When trained on the noisy data, the LFCC-LCNN model achieves an EER of 6.88\% and 29.67\% on the noisy seen and unseen test sets, respectively. When trained on the codec data, the LFCC-LCNN model achieves an EER of 1.45\% and 25.86\% on the codec seen and unseen test sets, respectively. In cases where only the data version is mismatched but the test set is seen, three models exhibit higher EERs. However, there are some exceptions, such as the LFCC-LCNN model trained on noisy data and tested on clean seen set, which achieves an EER of 3.43\%, and the RawNet2 model trained on noisy data and tested on codec seen set, which achieves an EER of 9.78\%. Another exception is the RawNet2 model trained on clean data and tested on codec seen set, which achieves an EER of 3.94\%. We speculate that certain features learned by the model are robust to audio format conversions, but further research is needed to understand the specific mechanism. Overall, in most cases,the detection models are not robust to noise and audio transcoding, especially on unseen data.

Table \ref{tab:fake_type} shows the EER of the model tested on different fake types. 
For testing with the "seen" type, all real audio samples from the seen test set and fake audio samples of the current type will be utilized. Similarly, for testing with the "unseen" type, all real audio samples from the unseen test set and fake audio samples of the current type will be utilized.
For the same fake type, there are differences in the test EERs of different systems. For example, when trained and tested on clean data, LFCC-LCNN achieves an EER metric of 7.73\% tested on F10, while LFCC-GMM achieves an EER of 12.46\% and RawNet2 achieves an EER of 46.98\%. LFCC-GMM and LFCC-LCNN have similar performance on F02 (EER of 3.30\% and 2.34\%, respectively) when training with clean data, but RawNet2 achieves an EER of 19.07\%. For all systems, F12 is the most difficult to detect of all fake types. 
The results of testing the fake type F12 in all cases show that the traditional model LFCC-GMM performs best. 
This suggests that GMM model still has a valuable place in fake audio detection, particularly when dealing with unseen data and data type mismatches. 
In a matched situation, neural network models can often outperform GMM model owning to their ability to learn complex representations.

We further compare the performance under different SNRs in Table \ref{tab:fake_snr}. LCNN performs best in the case where the test data matches the training data (the model trained with noisy data and tested on noisy seen data). In other cases, all three systems tested poorly. The worst case is LFCC-LCNN trained with clean data and tested on noisy unseen data at 0dB. Most of the results show that EER goes lower as SNR increases, which means noisy audio with high SNR are eaiser to detect and noisy audio with low SNR are difficult to detect. 

As for the performance under different transcoding operations, Table \ref{tab:fake_codec} demonstrates the results. It can be observed that RawNet2 has the potential to deal with different codecs whereas the LFCC-LCNN model trained on clean data is not robust under transcoding conditions. For seen codecs, LFCC-LCNN model performs similarly EERs in different codecs. The EERs of LFCC-LCNN model tested in unseen codecs are also close to each other. Other models exhibit the same pattern.

\subsection{Fake Algorithm Recognition}
Fake algorithm recognition is to classify fake audio into its correct category. In many application scenarios, e.g. judicial forensics by Court, not only do we care about the authenticity of the audio itself, but also need to know what model or algorithm generates it.
For example, if an audio recording is submitted as evidence to the Court, we need to verify whether the audio is authentic. If the audio is detected as fake audio, further fake algorithm recognition can provide an explanation. Besides, fake algorithm recognition can also increase the explainability of results in normal fake audio detection. Now, a few studies\cite{yan2022initial, yan2022system, bartusiak2022transformer} have focused on this area. Because the research has just started, recognizing seen types is the basic requirement, so we only select the seen test set for the recognition test.

Two baseline systems selected for fake algorithm recognition are:

\begin{enumerate}

    \item \textbf{LFCC-x-vector}: This system operates upon LFCC features with time delay neural networks (TDNN). X-vector \cite{snyder2018x} extracted from TDNN is robust embeddings for speaker recognition. The frame length and shift of LFCC are set to 25ms and 20ms respectively. The architecture of TDNN refers to \cite{peddinti2015time}.
    
    \item \textbf{LFCC-LCNN}: The setting of the LFCC feature is the same as in the LFCC-x-vector system. The LCNN model is the same as in fake audio detection task.

\end{enumerate}

\begin{table}[tb]
    \caption{The overall $F_1$-score of fake algorithm recognition.}
    \label{tab:f1_all}
    \centering
    \begin{tabular}{ cc|c|c|c }
    \toprule
    Train Set & Model & \makecell[c]{Clean \\ Seen Test}  
    & \makecell[c]{Noisy \\ Seen Test} &\makecell[c]{Codec \\ Seen Test} \\
    \midrule
    \multirow{3}*{Clean} & LFCC-X-vector &    93.72 &53.20  &30.28   \\
    ~ & LFCC-LCNN   &   97.26 &48.86  &64.30        \\
    \midrule
   \multirow{3}*{Noisy} & 
   LFCC-X-vector &   95.24 & 94.13  &39.09   \\
    ~ & 
    LFCC-LCNN     &   96.74 &93.63  &44.07       \\
    \midrule
   \multirow{3}*{Codec} & 
   LFCC-X-vector &   76.89 &	31.56	&98.02
    \\
    ~ & 
    LFCC-LCNN     &  97.31	&45.29	&95.04
      \\
    \bottomrule
  \end{tabular}
\end{table}

\begin{table*}[tb]

    \caption{The $F_1$-score (\%) of the model trained with different training data sets on different fake types.}
    \label{tab:alg_fake}
    \centering
    \setlength\tabcolsep{5.5pt}
\begin{tabular}{cc|c|c|c|c|c|c|c|c}
\toprule 
\multirow{2}{*}{Train Set} & \multirow{2}{*}{Model} & \multicolumn{8}{c}{Clean Test} \\
\cline{3-10} 
~& ~& F01 & F02& F03  & F04& F05 & F06 &F07  & F08  \\ 
\midrule
\multirow{2}{*}{Clean}     & LFCC-X-vector          & 99.91& 99.54 & 100.00 & 99.67 &81.89 & 99.76 & 70.72 & 98.31 \\
& LFCC-LCNN              & 100.00&100.00& 100.00 & 99.96 &92.02 &98.16 & 88.09 & 99.86  \\ 
\midrule
\multirow{2}{*}{Noisy}     & LFCC-X-vector   & 93.87 & 95.43  &93.95  & 97.97 & 93.55 & 98.40 & 92.51 & 96.21 
 \\
~ & LFCC-LCNN  & 98.84  & 99.63 &99.73 &98.46 & 89.81 &  98.88  &  90.14  & 98.44  \\ 
\midrule
\multirow{2}{*}{Codec}     & 
 LFCC-X-vector   & 93.12	 &99.96	 &100.00 &	80.50 &	72.75 &	69.05 &	27.20 &	72.56
 \\
~ &  
LFCC-LCNN  & 99.93 &	98.18	 &100.00 &	98.33	 &94.87 &	96.09 &	93.04	 &98.02
  \\ 
\midrule

\multirow{2}{*}{} & \multirow{2}{*}{} & \multicolumn{8}{c}{Noisy Test}  \\ 
\cline{3-10} 
~ & ~ & \multicolumn{1}{c|}{F01}   & \multicolumn{1}{c|}{F02}   & \multicolumn{1}{c|}{F03}   & \multicolumn{1}{c|}{F04}   & \multicolumn{1}{c|}{F05}   & \multicolumn{1}{c|}{F06}   & \multicolumn{1}{c|}{F07}   & F08   \\ 
\midrule
\multirow{2}{*}{Clean}     & LFCC-X-vector          & \multicolumn{1}{c|}{53.29} & \multicolumn{1}{c|}{75.13} & \multicolumn{1}{c|}{14.51} & \multicolumn{1}{c|}{61.33} & \multicolumn{1}{c|}{56.82} & \multicolumn{1}{c|}{74.55} & \multicolumn{1}{c|}{24.32} & 62.26\\
~ & LFCC-LCNN              & \multicolumn{1}{c|}{40.09} & \multicolumn{1}{c|}{76.18} & \multicolumn{1}{c|}{19.04} & \multicolumn{1}{c|}{43.15} & \multicolumn{1}{c|}{57.25} & \multicolumn{1}{c|}{64.05} & \multicolumn{1}{c|}{32.25} & 54.66 \\ 
\midrule
\multirow{2}{*}{Noisy}     & LFCC-X-vector          & \multicolumn{1}{c|}{91.57} & \multicolumn{1}{c|}{95.41} & \multicolumn{1}{c|}{95.51} & \multicolumn{1}{c|}{94.69} & \multicolumn{1}{c|}{92.33} & \multicolumn{1}{c|}{95.05} & \multicolumn{1}{c|}{90.34} & 97.91 \\
~& LFCC-LCNN              & \multicolumn{1}{c|}{97.80} & \multicolumn{1}{c|}{99.02} & \multicolumn{1}{c|}{99.32} & \multicolumn{1}{c|}{96.12} & \multicolumn{1}{c|}{85.70} & \multicolumn{1}{c|}{96.98} & \multicolumn{1}{c|}{81.30} & 91.99\\ 
\midrule
\multirow{2}{*}{Codec}     & 
 LFCC-X-vector   & 18.62	&63.81	&41.90&	35.94&	42.39&	9.77	&15.44&	19.59
 \\
~ & 
LFCC-LCNN  & 28.54	&66.60	&64.75&	39.17	&40.99	&38.57&	49.00	&31.53
  \\ 
\midrule
\multirow{2}{*}{Train Set} & \multirow{2}{*}{Model} & \multicolumn{8}{c}{Codec Test} \\
\cline{3-10} 
~& ~& F01 & F02& F03  & F04& F05 & F06 &F07  & F08  \\ 
\midrule
\multirow{2}{*}{Clean}     & 
LFCC-X-vector          & 37.91&	66.71	&66.65	&13.78	&30.02	&7.14	&4.27	&15.77
 \\
& 
LFCC-LCNN              &  78.71	&59.25	&80.63&	70.46	&84.81	&46.67	&20.37	&73.49
 \\ 
\midrule
\multirow{2}{*}{Noisy}     & 
LFCC-X-vector   &  47.46	&71.25	&44.51&	21.15	&37.54	&5.85	&37.10	&47.88
 \\
~ & 
LFCC-LCNN  &   68.98	&64.26&	60.26&	26.52&	54.14	&1.52	&15.20	&61.69
\\ 
\midrule
\multirow{2}{*}{Codec}     & 
LFCC-X-vector   & 98.49&	99.13	&99.79	&98.95	&98.70	&98.02	&94.90&	96.18
 \\
~ & 
LFCC-LCNN  &99.96	&100.00	&99.97	&99.89	&98.91	&99.94	&76.76	&84.90
  \\ 
\bottomrule
\end{tabular}

\end{table*}

\begin{table*}[tb]
    \caption{The $F_1$-score (\%) of the model trained with different training data sets at different SNRs.}
    \label{tab:alg_snr}
    \centering
     \setlength\tabcolsep{8pt}
\begin{tabular}{cc|c|c|c|c|c}
\hline
Train Set              & Model         & 0dB     & 5dB     & 10dB    & 15dB    & 20dB    \\ 
\toprule
\multirow{2}{*}{Clean} & LFCC-X-vector & 42.29 & 45.67 & 50.91 & 57.05 & 62.28  \\
                       & LFCC-LCNN     & 33.99 & 36.03 & 45.98 & 53.90 & 61.96 \\ 
                       \midrule
\multirow{2}{*}{Noisy} & LFCC-X-vector & 90.11 & 93.71 & 94.33 & 95.54 & 95.91  \\
                       & LFCC-LCNN     & 90.85 & 92.58 & 93.68 & 94.57 & 95.25  \\ 
                       \bottomrule
\end{tabular}

\end{table*}

Table \ref{tab:f1_all} shows the results tested on seen test set. 
For clean audio, the LFCC-LCNN model achieves better performance across all cases of the training set used, achieving an F1-score of 97.26\% when trained on the clean set, 96.74\% when trained on the noisy set, and 76.89\% when trained on the codec set. But it's hard to say which model is good at fake algorithm recognition under noisy and codec conditions. In scenarios where both the training and test sets are noisy or codec, LFCC-x-vector model has some advantages over LFCC-LCNN model. When dealing with unmatched data  (model trained and tested on different verison data), both the LFCC-x-vector and LFCC-LCNN models exhibit a significant drop in their F1-score (LFCC-x-vector achieves an F1-score of 39.09\% when trained on noisy set and tested on codec set, and it achieves an F1-score of 31.56\% when trained on codec set and tested on noisy set. The performance of LFCC-LCNN follows a similar pattern.). 
We can conclude that for cases of data mismatch, recognizing fake algorithms under noisy or transcoding conditions is significantly more challenging than under clean conditions. An interesting finding is that the model trained on noisy data is robust to clean data and only LFCC-LCNN model trained on codec data is robust to clean data. We speculate that LFCC-LCNN model is more robust to audio transcoding.

Table \ref{tab:alg_fake} demonstrates the detailed result tested on each fake type. Under clean matched condition (model trained on clean data and tested on clean data), type F01-F04 can be distinguished by the models with great accuracy, almost 100 \% F1-score. But the F1-score of F05 and F07 is relatively low. 
For the clean test condition, F07 is a challenging type to recognize. But interestingly, the performance (F1-score on F07) of the model trained with noisy set is better than the model trained with clean set.
For the noisy test set, F03 is the most difficult for models trained with clean data, and the F1-score is less than 20\%. 
For the codec test set, F07 is the most difficult type for the model trained with clean data.
In all types of fake audio, models trained with noisy data perform better than models trained with clean data when tested on noisy data. 


We further compare the recognition performance under different SNRs in Table \ref{tab:alg_snr}. The results show that the F1-score value of each model increases gradually with the increase of SNR. In the case of SNR20, both X-vector and LCNN models achieve the best performance. Recognizing the fake algorithm is more difficult under lower SNR than under higher SNR, which is consistent with fake audio detection task.

\section{Discussions}
\begin{figure}[t]
  \centering
    \subfigure[]{
    \includegraphics[width=0.95\linewidth]{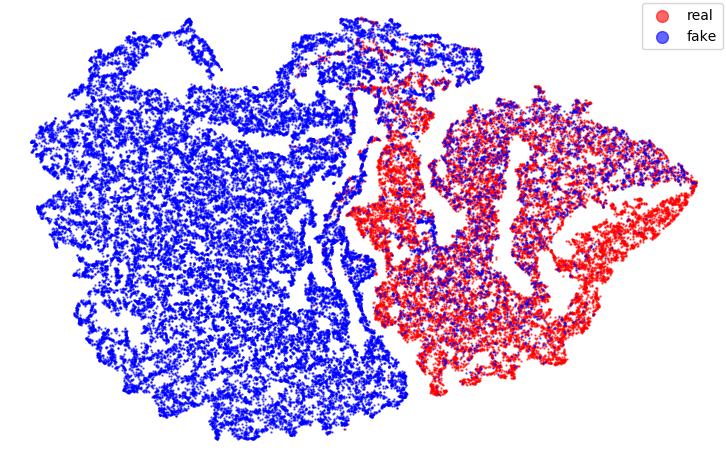} }
    \subfigure[]{
    \includegraphics[width=0.95\linewidth]{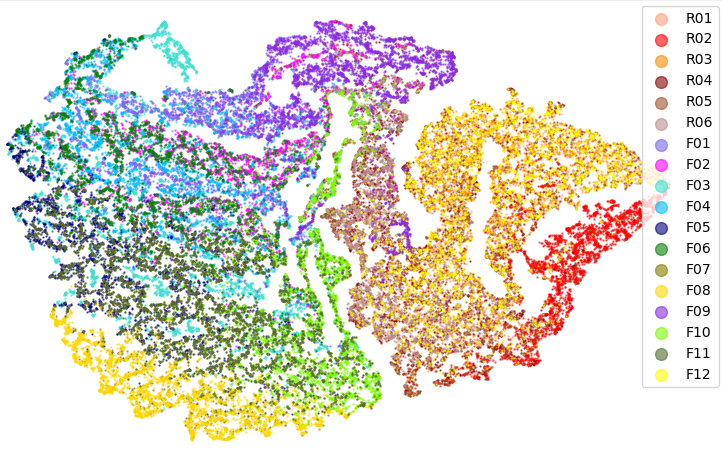}}
    
  \caption{Visualization of real and fake audio data of CFAD clean test set. Color blue in Fig(a) represents fake audio. Color red in Fig(a) represents real audio. Different colors in Fig(b) represent different real or fake types.}
  \label{fig:embedding}

\end{figure}

We plot t-SNE visualization of the embeddings of audio in clean test set extracted by the LFCC-LCNN system. LCNN is trained for fake audio detection task.
In Figure \ref{fig:embedding}(a), fake audio (blue) and real audio (red) are divided into two parts. The audio near the class boundary is easily misclassified. Further, fake and real categories are divided into several sub-categories. Each sub-category represents a generation method or resource and can be identified by
the colors shown in Figure \ref{fig:embedding}(b). We find that audio of the same sub-category has a relatively similar distribution, and audios of different sub-categories are different. 
While some of the speech synthesis systems are well separated from the real audio, fake audio from F02, F09, F10, F11, and F12 overlap
with the real audio. For example, F09 and F10 are located near real audio, resulting in some data points being mixed up. More seriously, the F12 falls almost entirely in the real audio area. 
We believe this is because the data is unseen during training, causing the model to make mistakes in its feature representation.
It can be seen from the experimental results that the EER of F12 is the highest (70.62\%), and it's the most difficult for the model to detect fake type F12.
Besides, F06, F10, and F11 are colored in different greens. They overlap with each other because they share the same vocoder HifiGAN.
The distribution of real audio from different corpora is different.
R05 and R06 are located near the class boundary, while R01 and R02 can be well separated from fake audio. This is still a result of whether the data has been seen during training. This again illustrates the necessity of considering the diversity of real speech resources to enhance the generalization of the detection model.


    

\section{Future Directions} 

We have designed an initial Chinese public dataset under additive noise conditions for fake audio detection and fake algorithm recognition. There are still some limitations that are suggested to be potential research directions in the future.

\textbf{Simulating utterance under more acoustic conditions}:
 The simulated noisy genuine and fake utterances of the current CFAD dataset are under additive noise conditions. However, there are more complex noise scenarios in real life. More noises are utilized to generate noisy utterances, such as convolutional noises.

\textbf{Generating noisy audio with matched linguistic content}:
The noisy utterances of our CFAD dataset are simulated by randomly adding noise signals to clean utterances. The linguistic content and the noise of the audio may exist mismatched. To make noisy data more reasonable in practical applications, we need to consider the match between the linguistic content and the noise.

\textbf{Collecting noisy audio under realistic conditions}: The noisy utterances of the CFAD dataset are simulated data. Such simulations do not quite match the real and fake utterances collected in real conditions. In order to evaluate the robustness and generation of fake audio detection methods in practical applications, the noisy genuine and fake utterances are suggested to collect under realistic environmental conditions.

\textbf{More diverse audio codecs}:
 The codec version of CFAD dataset contains 6 kinds of audio codecs. But codecs is more diverse and complex in real-life scenarios. Some audio may undergo successive transcoding operations of two different codecs. Besides, the configurations such as variable bit rate can be taken into account.

\textbf{More diverse real and fake audio types}: 
The CFAD dataset contains 6 kinds of real utterances and 12 sorts of fake attacks. But the audio is more diverse and complex in real-life scenarios. The fake audio generation methods based on voice conversion (VC) systems and phase variations and speaker/pitch/energy editing are suggested to consider. Besides, more datasets  need to be incorporated to traine TTS/VC systems.
It is crucial to take more diverse types of real and fake audio into consideration so that make the dataset is more appropriate for real scenarios.

\textbf{Generalization of detection methods}:
The work here aims to provide benchmark results on the CFAD dataset for future research. Better methods would be proposed to make the detection models generalize well to unknown types and mismatch conditions, such as continual learning, etc.

\textbf{Recognizing unseen fake algorithms}: 
The current work here only provides benchmark results for recognizing seen fake algorithms. In fact, there are many new types of fake utterances in real applications. So, models need to recognize unseen fake attacks.

\section{Conclusions}

The generalization of fake audio detection models is a significant challenge for current detection methods, and there is a lack of standard publicly available Chinese dataset to support relevant research. This paper presents the first public Chinese standard dataset for fake audio detection under complex conditions. It meets the need to detect fake audio accompanied by a variety of background noises and processed using different codec in real-life scenarios and further recognize the exact types of fake audio. The design process and baseline results for fake audio detection and fake algorithm recognition are reported. Future work has been mentioned in the previous section.


\section{Acknowledgments}
This work is supported by the National Key Research
and Development Plan of China (No.2020AAA0140003),
the National Natural Science Foundation of China (NSFC)
(No.61901473, No.62101553, No.61831022).

\newpage



\bibliographystyle{IEEEtran}
\bibliography{tifs202209}


\vspace{-20pt} 
\begin{IEEEbiography}[{\includegraphics[width=1in,height=1.25in,clip,keepaspectratio]{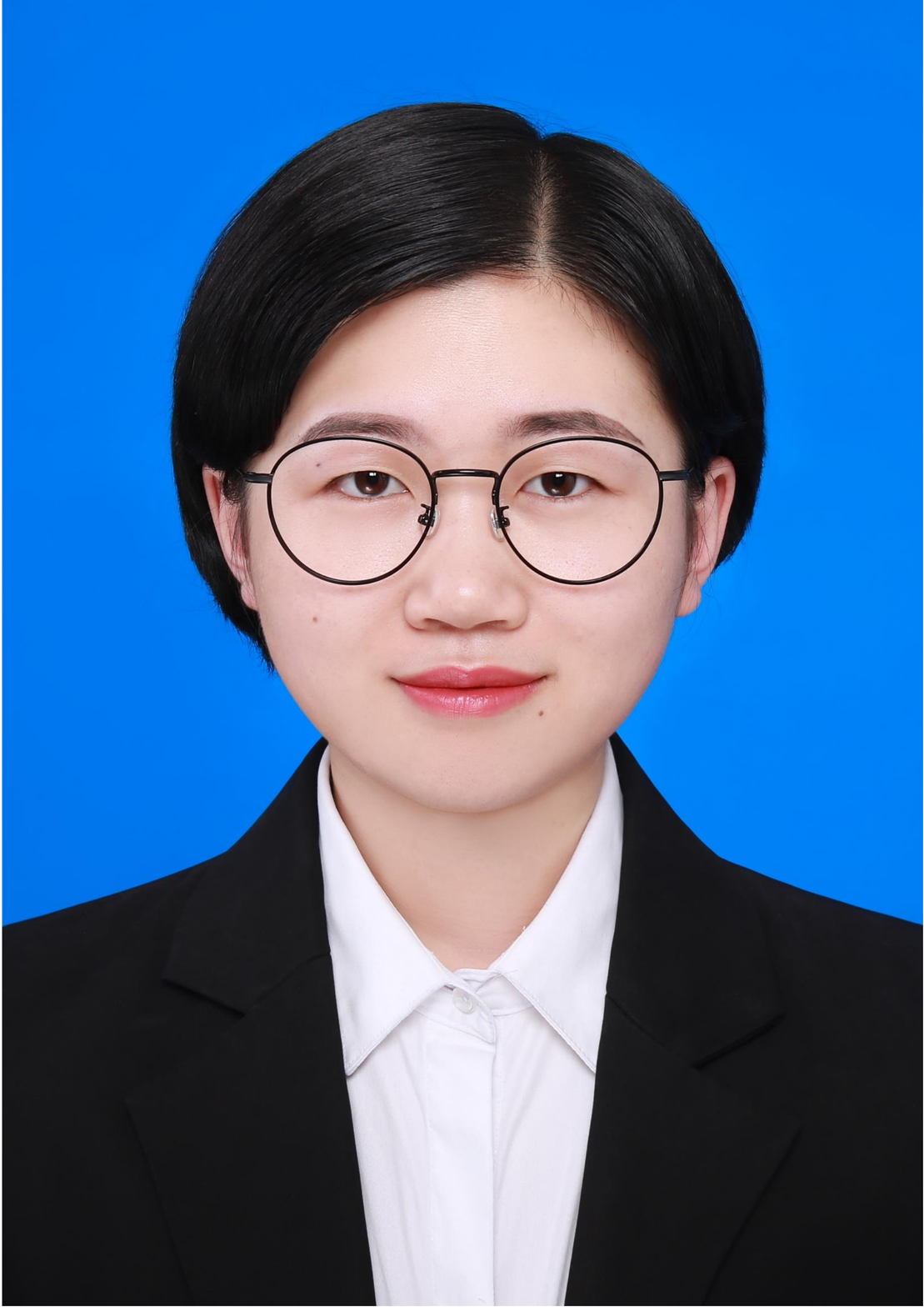}}]
{Haoxin Ma}
received the B.S. degree from Southeast University, Nanjing, China, in 2019, and the M.S. degree from the University of Chinese Academy of Sciences, Beijing, China, in 2022. Her current research interest is deepfake audio detection.
\end{IEEEbiography}

\vspace{-20pt} 
\begin{IEEEbiography}[{\includegraphics[width=1in,height=1.25in,clip,keepaspectratio]{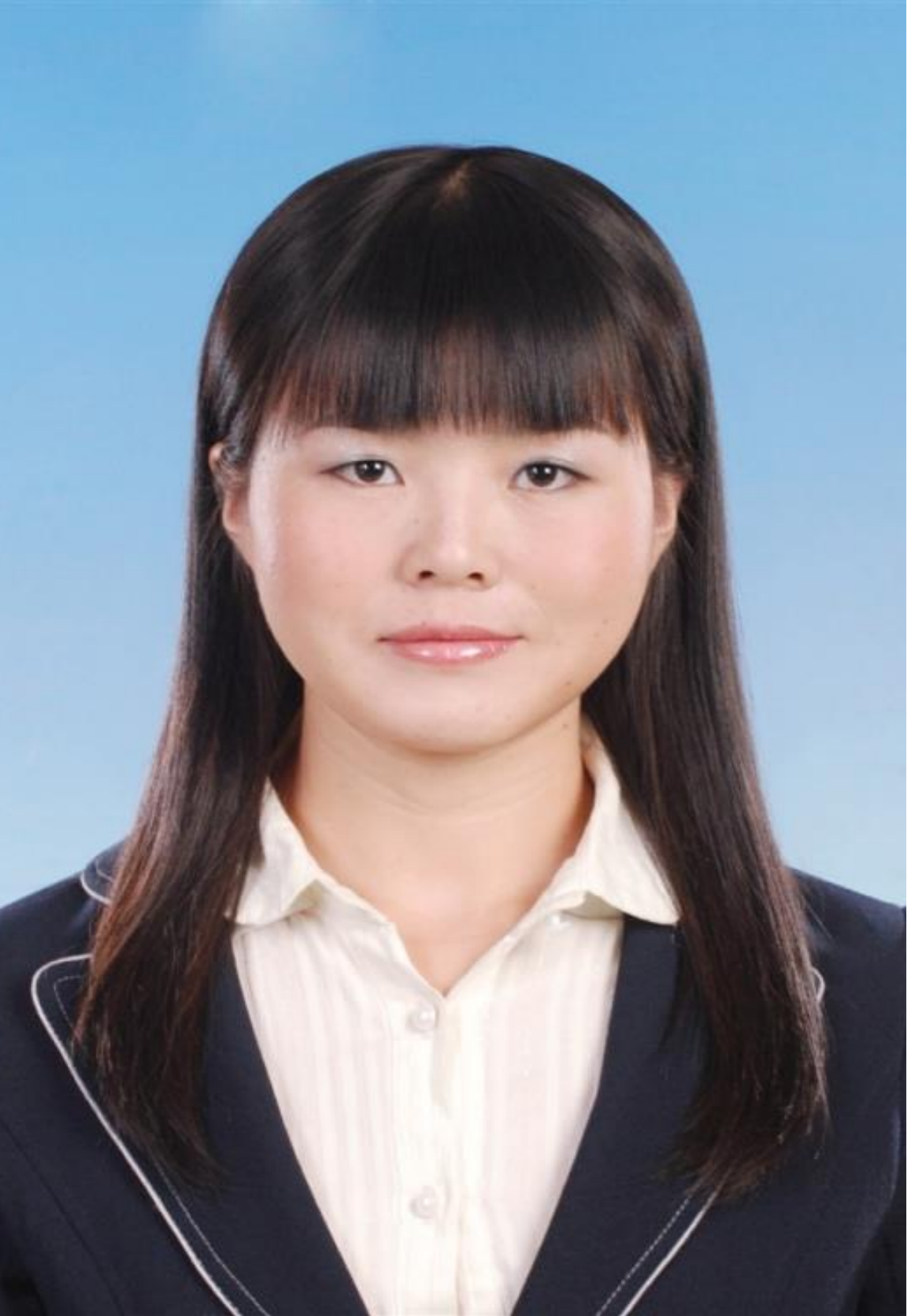}}]{Jiangyan Yi}
 received the Ph.D. degree from the University of Chinese Academy of Sciences, Beijing, China, in 2018, and the M.A. degree from the Graduate School of Chinese Academy of Social Sciences, Beijing, China, in 2010. She was a Senior R\&D Engineer with Alibaba Group during 2011 to 2014. She is currently an Associate Professor with the National Laboratory of Pattern Recognition, Institute of Automation, Chinese Academy of Sciences, Beijing, China. Her current research interests include speech signal processing, speech recognition and synthesis, fake audio detection, audio forensics and transfer learning.
\end{IEEEbiography}
\vspace{-30pt} 
\begin{IEEEbiography}[{\includegraphics[width=1in,height=1.25in,clip,keepaspectratio]{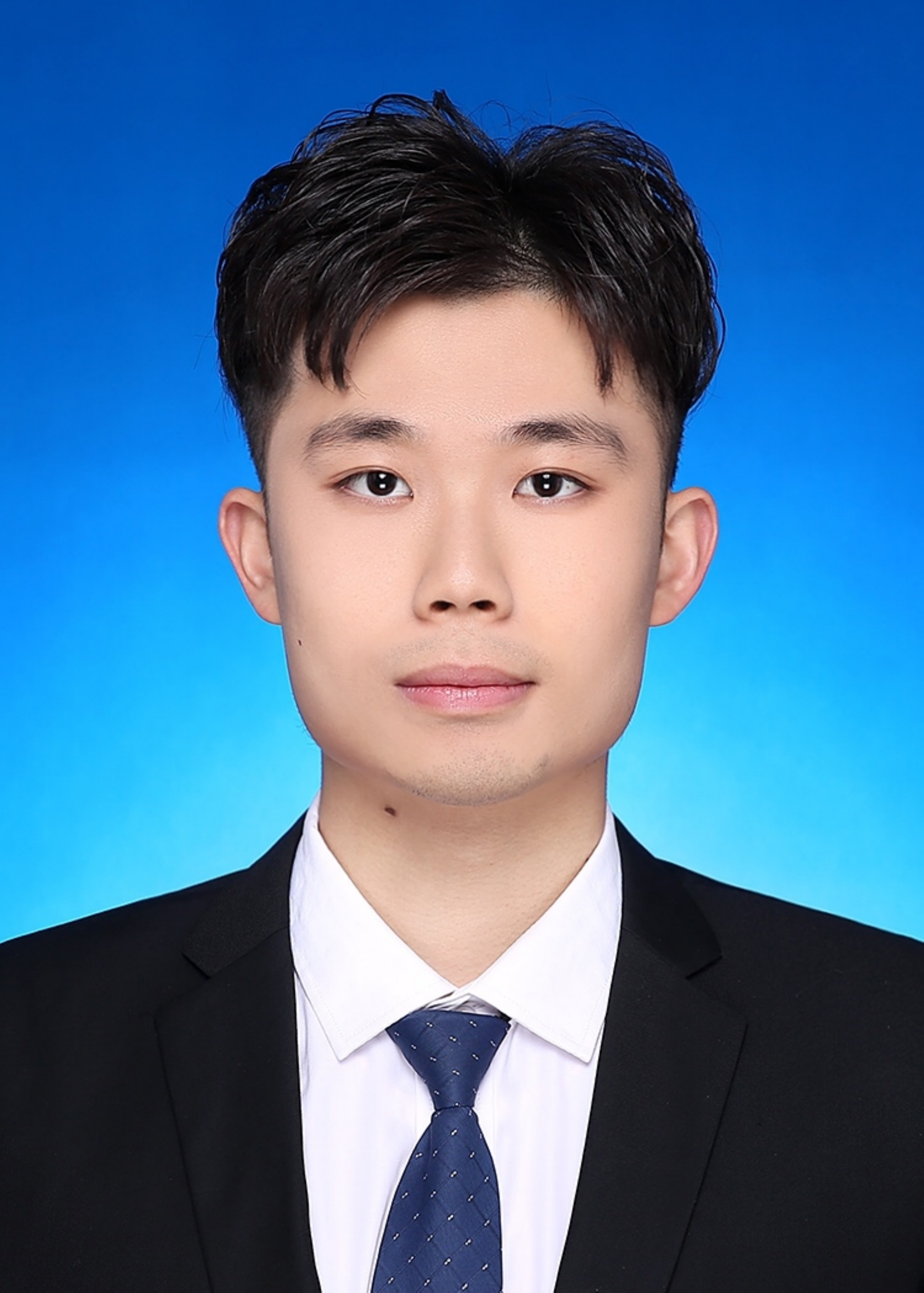}}]{Chenglong Wang}
 received the B.S. degree from Hefei University of Technology, Anhui, China, in 2018. He is currently working toward the Ph.D. degree with the University of Science and Technology of China, Anhui, China. His current research interests include fake audio detection, speaker verification and identification.
\end{IEEEbiography}
\vspace{-40pt}
\begin{IEEEbiography}[{\includegraphics[width=1in,height=1.25in,clip,keepaspectratio]{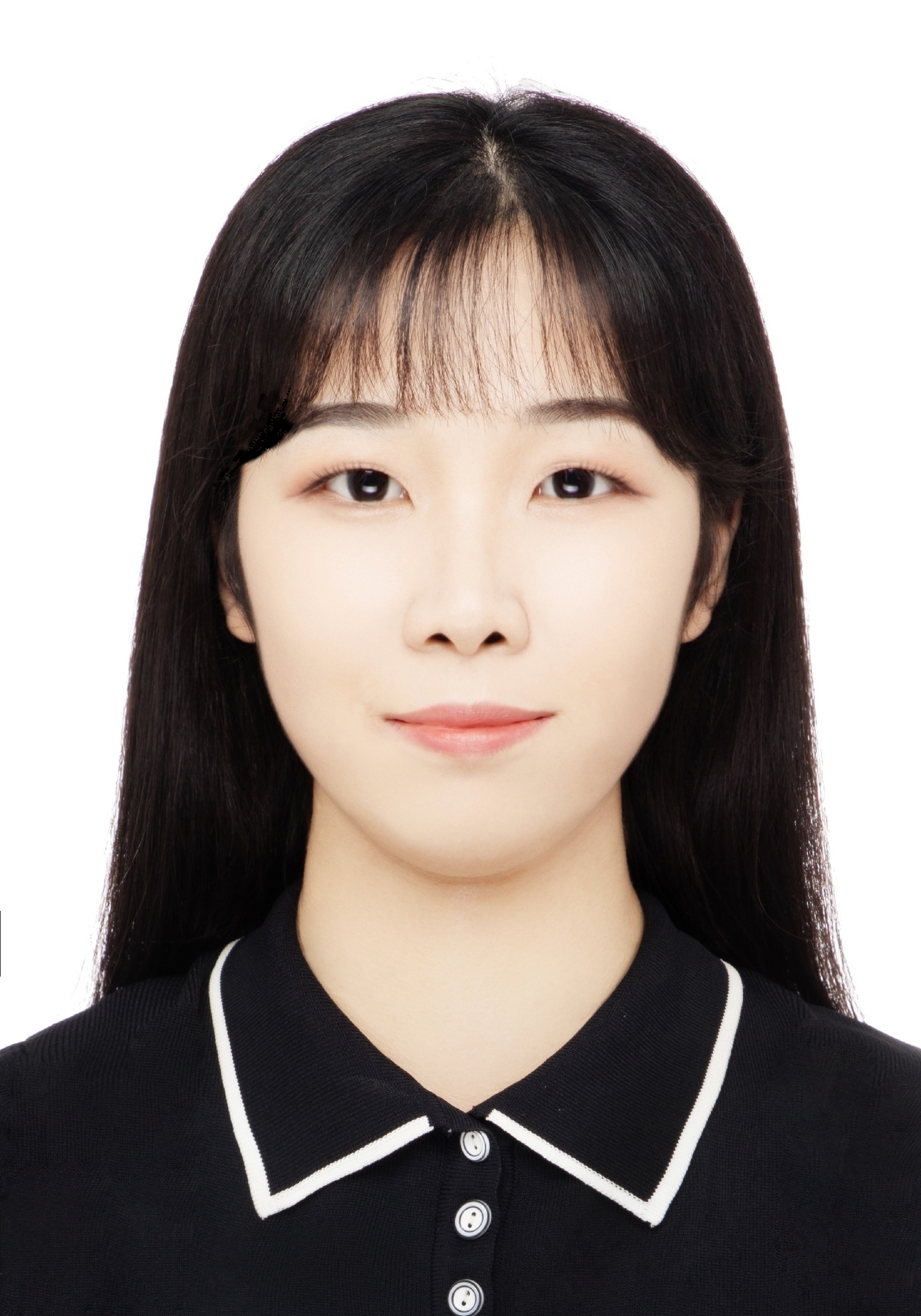}}]{Xinrui Yan}
received the B.S. degree from Northeastern University in China in 2021. She is currently pursuing her M.S. degree at the University of Chinese Academy of Sciences in Beijing, China. Her current research interest is audio fake forensics.
\end{IEEEbiography}
\vspace{-40pt}
\begin{IEEEbiography}[{\includegraphics[width=1in,height=1.25in,clip,keepaspectratio]{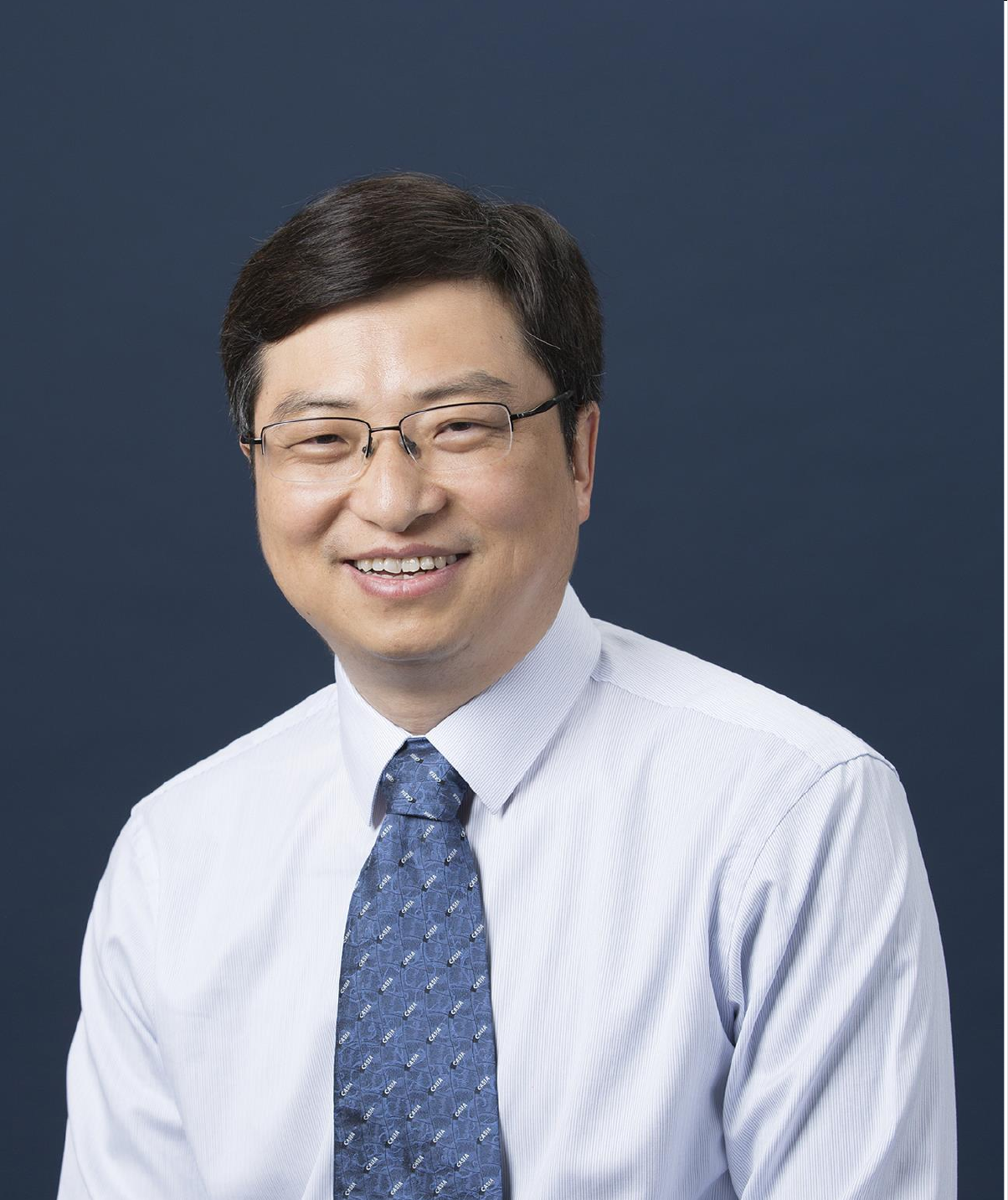}}]{Jianhua Tao}
 received his Ph.D. degree from Tsinghua University, Beijing, China, in 2001, and the M.S. degree from Nanjing University, Nanjing, China, in 1996. He is currently a Professor with NLPR, Institute of Automation, Chinese Academy of Sciences, Beijing, China. He has authored or coauthored more than eighty papers on major journals and proceedings including the IEEE TRANSACTIONS ON AUDIO, SPEECH, AND LANGUAGE PROCESSING. His current research interests include speech signal processing, speech recognition and synthesis, human computer interaction, multimedia information processing, and pattern recognition.
\end{IEEEbiography}
\vspace{-30pt}
\begin{IEEEbiography}[{\includegraphics[width=1in,height=1.25in,clip,keepaspectratio]{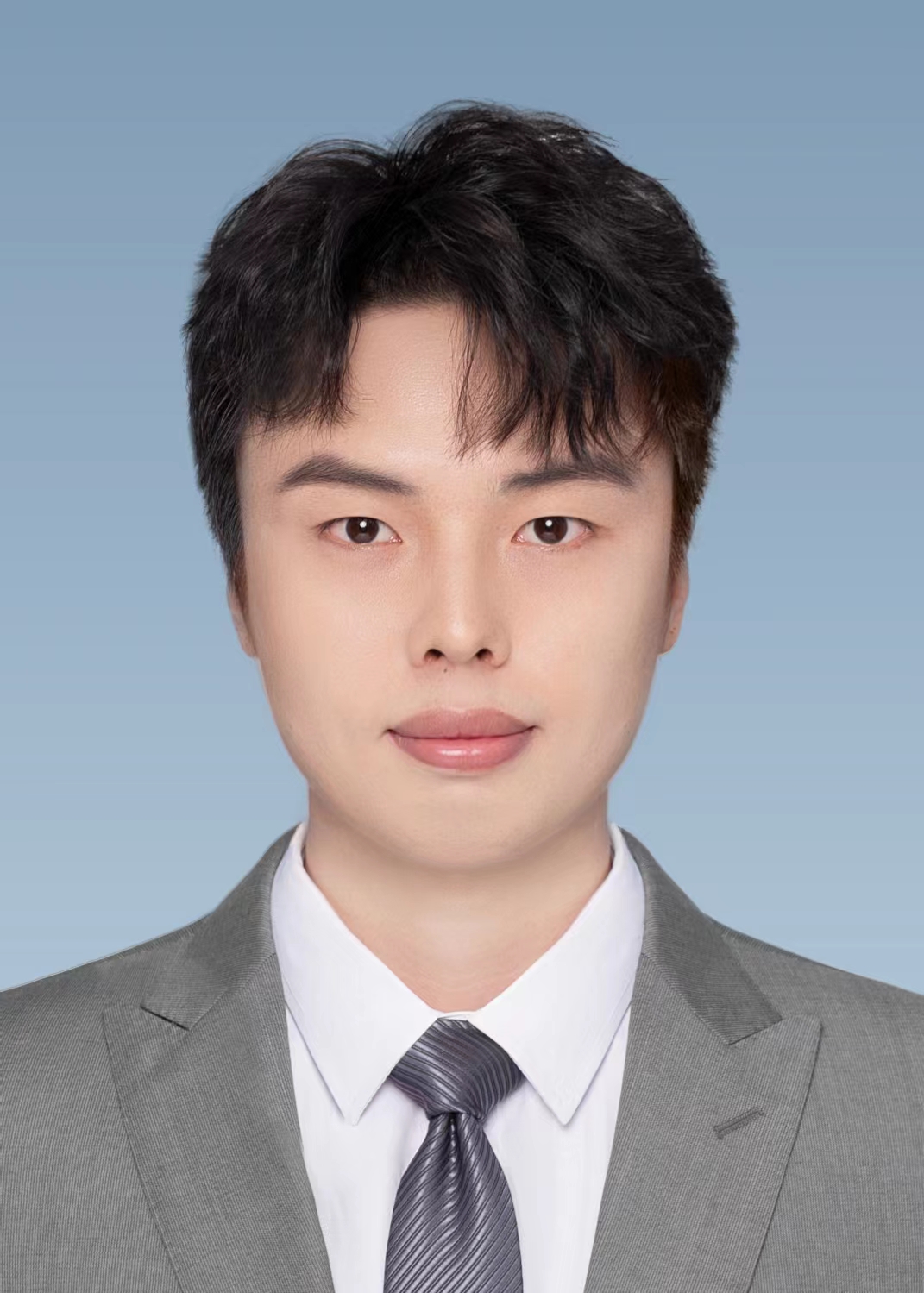}}]{Tao Wang}
received the B.E. degree from the
Department of  Control Science and Engineering, Shandong University (SDU), Jinan, China, in 2018. He is currently working
toward the Ph.D. degree with the National Laboratory
of Pattern Recognition, Institute of Automation (NLPR), Chinese
Academy of Sciences (CASIA), Beijing, China. His current
research interests include speech synthesis, voice conversion, speech editing, machine learning, and transfer learning.
\end{IEEEbiography}
\vspace{-30pt}
\begin{IEEEbiography}[{\includegraphics[width=1in,height=1.25in,clip,keepaspectratio]{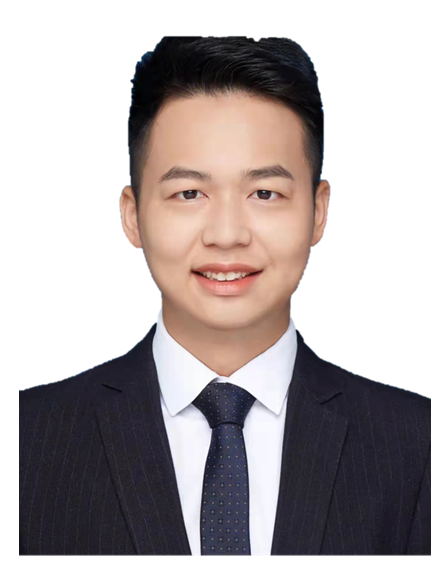}}]{Shiming Wang}
 received the B.S. degree from  Northwestern Polytechnical University,Xi‘an, China, in 201. He is currently working toward the Ph.D. degree with the University of Science and Technology of China, Anhui, China. His current research interests include speech synthesis, self-supervised pretrain model of audio.

\end{IEEEbiography}

\vspace{-25pt}
\begin{IEEEbiography}[{\includegraphics[width=1in,height=1.25in,clip,keepaspectratio]{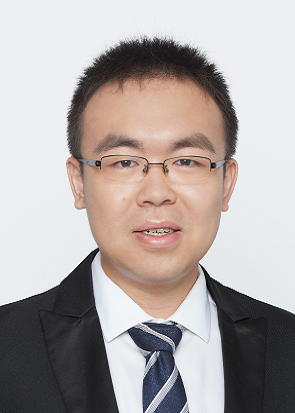}}]{Ruibo Fu}
 is an assistant professor in the National
 Laboratory of Pattern Recognition, Institute of Automation, Chinese Academy
 of Sciences, Beijing. He obtained B.E. from Beijing University of Aeronautics and Astronautics in 2015 and Ph.D. from Institute of Automation, Chinese Academy of Sciences in 2020. His research interest is speech synthesis and transfer learning.
\end{IEEEbiography}


\end{document}